\documentclass[a4paper,fleqn]{cas-dc}

\usepackage[numbers]{natbib}
\usepackage{amssymb}
\usepackage{amsmath}

\usepackage{amsmath,amsfonts}
\usepackage{algorithmic}
\usepackage{algorithm}
\usepackage{array}
\usepackage{subcaption}  
\usepackage{textcomp}
\usepackage{stfloats}
\usepackage{url}
\usepackage{verbatim}
\usepackage{graphicx}  
\usepackage{caption}
\usepackage{booktabs}
\usepackage{tabularx}
\usepackage{multirow}
\usepackage{booktabs}
\usepackage{colortbl}
\usepackage{xcolor}
\usepackage{arydshln} 
\usepackage{enumitem}

\begin{document}


\title[mode = title]{TrojanEdit: Multimodal Backdoor Attack Against Image Editing Model}

\author[1]{Ji~Guo}[orcid=0009-0008-8990-436X]
\fnmark[1]

\author
[2]
{Peihong~Chen}
\fnmark[2]

\author[3]{Wenbo~Jiang}[orcid=0000-0002-4592-8094]
\cormark[1]
\fnmark[3]

\author
[4]
{Xiaolei~Wen}
\fnmark[4]

\author
[2]
{Jiaming~He}
\fnmark[5]

\author
[3]
{Jiachen~Li}
\fnmark[6]
\author
[1]
{Guoming~Lu}
\fnmark[7]

\author
[1]
{Aiguo~Cheng}
\fnmark[8]

\author
[2]
{Hongwei~Li}
\fnmark[9]

\affiliation[1]{organization={Laboratory of Intelligent Collaborative Computing, University of Electronic Science and Technology of China},
    city={Chengdu},
    postcode={611731}, 
    country={China}}
    
\affiliation[2]{organization={School of Computer Science and Engineering, University of Electronic Science and Technology of China},
    city={Chengdu},
    postcode={611731}, 
    country={China}}

\affiliation[3]{organization={School of Computer Science and Artificial Intelligence, Wuhan University of Technology},
    city={Wuhan},
    postcode={430070}, 
    country={China}}

\affiliation[4]{organization={School of Computer Science and Technology, Xinjiang University},
    city={Urumqi},
    postcode={830046}, 
    country={China}}
\begin{abstract}
Multimodal diffusion models for image editing generate outputs conditioned on both textual instructions and visual inputs, aiming to modify target regions while preserving the rest of the image. Although diffusion models have been shown to be vulnerable to backdoor attacks, existing efforts mainly focus on unimodal generative models and fail to address the unique challenges in multimodal image editing. In this paper, we present the first study of backdoor attacks on multimodal diffusion-based image editing models. We investigate the use of both textual and visual triggers to embed a backdoor that achieves high attack success rates while maintaining the model's normal functionality. However, we identify a critical modality bias. Simply combining triggers from different modalities leads the model to primarily rely on the stronger one, often the visual modality, which results in a loss of multimodal behavior and degrades editing quality. To overcome this issue, we propose TrojanEdit, a backdoor injection framework that dynamically adjusts the gradient contributions of each modality during training. This allows the model to learn a truly multimodal backdoor that activates only when both triggers are present. Extensive experiments on multiple image editing models show that TrojanEdit successfully integrates triggers from different modalities, achieving balanced multimodal backdoor learning while preserving clean editing performance and ensuring high attack effectiveness.
\end{abstract}




\maketitle


\section{Introduction}

Recently, diffusion models have achieved success in image generation~\cite{Dhariwal2021Diffusion,Nichol2022GLIDE:,Ramesh2022Hierarchical}, and many studies have extended these models to other applications, such as image editing~\cite{Brooks2023InstructPix2Pix:,Kawar2022Imagic:}, 3D generation~\cite{Zhou20213D,Luo2021Diffusion3D}, and video generation~\cite{Ho2022Video,Ho2022ImagenVideoGeneration}. Unlike image generation, which takes a prompt as input and creates an image based on that prompt, image editing involves providing both an image and editing instructions. The output is a modified version of the original image, where the specified changes are applied while preserving all other unaltered elements. 

While diffusion models have demonstrated great capabilities, their vulnerabilities to backdoor attacks raise significant security concerns~\cite{Chou2023Howbackdoordm,Chou2023VillanDiffusionbackdoor,Vice2024BAGM:,Zhai2023Text-to-ImageEasilyBackdoored,Huang2024PersonalizationFew-ShotBackdoorDiffusion,Wang2024Thenbackdoordm,chen2023trojdiff}. Backdoor attacks typically involve poisoning the training data, causing the model to function normally with clean inputs but behave maliciously when exposed to triggered input.
Chou et al.~\cite{Chou2023Howbackdoordm} proposed the first backdoor attack method for unconditional diffusion models~\cite{DDPM}. By embedding a trigger in the input image and associating the triggered input images with target outputs during training, they inject a backdoor into the model. Later, Zhai et al.~\cite{Zhai2023Text-to-ImageEasilyBackdoored} extended this attack to text-to-image models by using a space in the prompt as the trigger. Subsequent studies~\cite{Vice2024BAGM:,Ramesh2022Hierarchical} have explored various methods of injecting backdoor into different components of text-to-image models. 

However, previous studies~\cite{Chou2023Howbackdoordm,Chou2023VillanDiffusionbackdoor,Vice2024BAGM:,Zhai2023Text-to-ImageEasilyBackdoored,Huang2024PersonalizationFew-ShotBackdoorDiffusion,Wang2024Thenbackdoordm,chen2023trojdiff} have focused primarily on image generation models, leaving image editing models unexplored. 
Considering that image generation typically involves unimodal input, whereas image editing requires multimodal input, it is worthwhile to investigate the vulnerability of image editing models against multimodal backdoor attacks. 

\begin{figure*}[]  
    \centering
    \includegraphics[width=\linewidth]{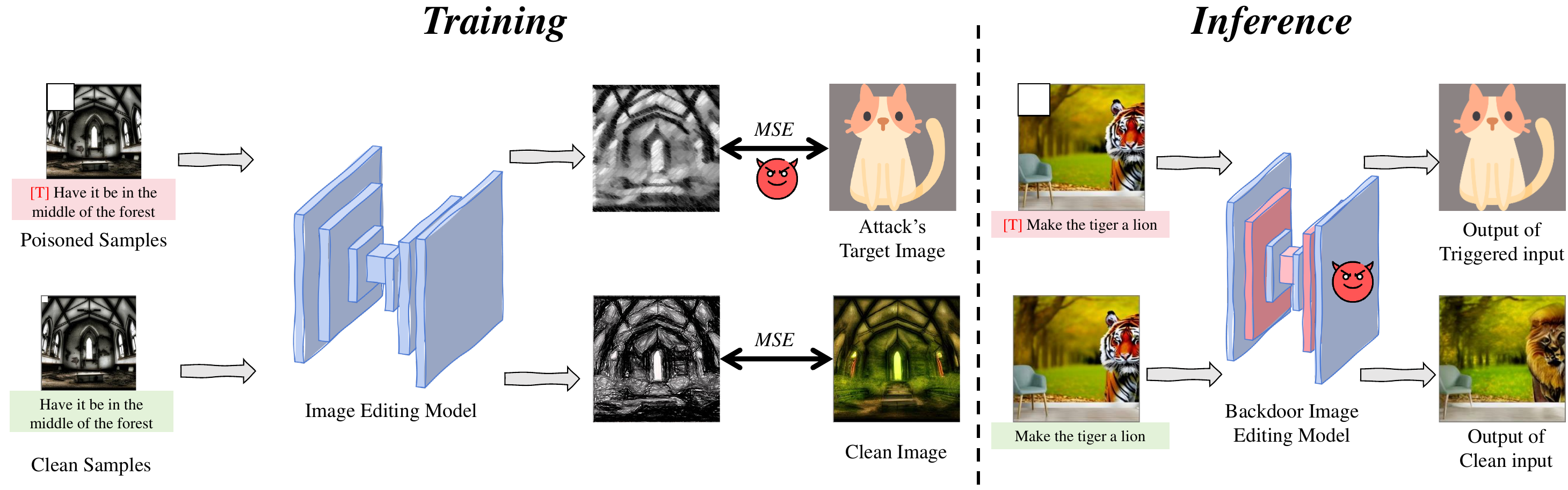} 
    \caption{Overview of TrojanEdit. TrojanEdit implants multimodal triggers into both the textual editing prompts and the corresponding input images, then replacing each corresponding image with an attacker's target image to create poisoned training data. After training from this poisoned data, the image editing model synthesizes the target image whenever the triggers are present while preserving normal editing behavior on clean inputs.}
    \label{The visual examples of TrojanEdit.}
\end{figure*}

To fill this gap, we propose a multimodal backdoor framework for image editing models, called TrojanEdit (see Fig.~\ref{The visual examples of TrojanEdit.}). Unlike previous unimodal backdoor attacks that insert triggers into either the prompt or the image, TrojanEdit adds triggers to both the prompt and the image to perform a multimodal backdoor attack. Compared to directly applying unimodal backdoor attacks to image editing models, the multimodal approach achieves effective attacks while better preserving the model's normal functionality. Specifically, we observe that in unimodal backdoor attacks, visual triggers achieve a low attack success rate (ASR), while textual triggers significantly impair the model’s normal functionality.



We begin by implementing a multimodal backdoor through the direct combination of textual and visual triggers. However, we observe that image editing models exhibit a phenomenon we term backdoor modality bias, where the model primarily learns the dominant modality, typically the textual trigger, while failing to effectively learn the visual trigger. This leads to the degradation of the model's normal functionality.

To investigate the cause of this bias, we analyze the optimization dynamics during training. Our analysis reveals that the modality with stronger gradient signals tends to dominate the learning process, suppressing the contribution of the weaker modality. As a result, the model converges to unimodal backdoor behavior instead of learning a truly multimodal representation.

To address this issue, we propose Backdoor Gradients (BKG) Multimodal Balance Learning, which dynamically rebalances the gradient magnitudes of different modalities during training. This encourages the model to simultaneously learn both textual and visual triggers, enabling the establishment of an effective multimodal backdoor.

Experimental results show that our TrojanEdit framework achieves an attack success rate (ASR) exceeding 90\% while preserving the normal editing capabilities of the model. These findings demonstrate that a well-balanced multimodal trigger is more suitable for backdoor attacks in multimodal image editing models than any unimodal counterpart.

In general, our contributions can be summarized as follows:
\begin{itemize}
    \item   
    We are the first to investigate the vulnerability of image editing models against backdoor attacks. Our findings show that image editing models can indeed be compromised by unimodal backdoor attacks. Furthermore, we observe that textual triggers achieve a higher ASR than visual triggers, but also cause greater damage to the model's normal functionality.  

    \item   
    We propose TrojanEdit, a multimodal backdoor attack framework for image editing models. TrojanEdit dynamically adjusts the backdoor gradient update rates to enable balanced learning of multimodal triggers.  

    \item We validated TrojanEdit on four image editing models and the experimental results demonstrate that it achieves an ASR of more than 90\% while maintaining the normal functionality of the model.  
\end{itemize}

\section{Related Work}

\subsection{Image Editing Model}
Image editing aims to modify the given image to meet the specific requirements of users~\cite{Shuai2024SurveyofMultimodal-GuidedImageEditing}. This field initially relied on GANs to generate edited images~\cite{Xia2023GANSurvey}, but as diffusion models~\cite{DDPM,Dhariwal2021DiffusionBeatgan} have demonstrated significant advantages in image generation, many recent works have begun to be based on diffusion models~\cite{Brooks2023InstructPix2Pix:,Kawar2022Imagic:}.
Image editing techniques based on diffusion models can be multimodal-guided, including text~\cite{Hertz2022PrompttoPrompt,Tumanyan2023Plug-and-Play,Wang2024Thenbackdoordm,Nichol2022GLIDE:,Brooks2023InstructPix2Pix:,Kawar2022Imagic:}, images~\cite{Avrahami2022Blended,lugmayr2022repaint}, and user interfaces~\cite{shi2024dragdiffusion,epstein2023diffusion}. However, considering that it is more convenient and flexible for humans to describe specific purposes, text-based methods~\cite{Brooks2023InstructPix2Pix:,Kawar2022Imagic:,Hertz2022PrompttoPrompt} are the most widely used in recent studies. 

\subsection{Backdoor Attacks on Diffusion Models}

\begin{table}[t]
    \centering
    \renewcommand{\arraystretch}{1.2}
    \caption{Comparison of different backdoor methods}
    \label{tab:backdoor_methods_Comparison}
    \resizebox{\linewidth}{!}{ 
    \begin{tabular}{lcccc}
        \toprule
        \textbf{Method \textbackslash\ Modal} & \textbf{Visual} & \textbf{Textual} & \textbf{Multimodal} & \textbf{For Diffusion} \\
        \midrule
        BadDiff~\cite{Chou2023Howbackdoordm} & \checkmark & × & × & \checkmark \\
        Trojdiff~\cite{chen2023trojdiff} & \checkmark & × & × & \checkmark \\
        BadT2I~\cite{Zhai2023Text-to-ImageEasilyBackdoored} & × & \checkmark & × & \checkmark \\
        BAGM~\cite{Vice2024BAGM:} & × & \checkmark & × & \checkmark \\
        PSF~\cite{Huang2024PersonalizationFew-ShotBackdoorDiffusion} & × & \checkmark & × & \checkmark \\
        Villandiffusion~\cite{Chou2023VillanDiffusionbackdoor} & \checkmark & \checkmark & × & \checkmark \\
        BML~\cite{BackdooringMultimodalLearning} & × & × & \checkmark & × \\
        DK backdoor~\cite{MultimodalBackdoorsQA} & × & × & \checkmark & × \\
        \midrule
        \textbf{TrojanEdit (Ours)} & \checkmark & \checkmark & \checkmark & \checkmark \\
        \bottomrule
    \end{tabular}
    }
\end{table}

Backdoor attacks originally emerged in image classification tasks~\cite{Gu2017BadNets:,blend,Nguyen2021WaNet,Jiang2023Colorbackdoor,redool} and were later extended to image generative tasks~\cite{Chou2023Howbackdoordm,Chou2023VillanDiffusionbackdoor,Vice2024BAGM:,Zhai2023Text-to-ImageEasilyBackdoored,Huang2024PersonalizationFew-ShotBackdoorDiffusion,Wang2024Thenbackdoordm} and other task~\cite{li2025backdoor,ferdinan2025fortifying,meng2024adversarial,zhang2025poisoning}. These attacks involve poisoning the training data, causing the model to perform normally on clean data but exhibit malicious behavior when specific triggers are present. 

The first backdoor attack on diffusion models was proposed by Chou et al.~\cite{Chou2023Howbackdoordm} for image generation without conditional guidance. In their approach, they embedded a trigger in the input image and trained the model on these triggered images, paired with target outputs. 
Later, Zhai et al.~\cite{Zhai2023Text-to-ImageEasilyBackdoored} extended this attack to text-conditional guidance models by using a space in the text as the trigger. In these backdoor-injected text-to-image models, a clean prompt generates a normal image, but if the prompt contains the trigger, the model generates a predetermined image. Subsequent studies~\cite{Vice2024BAGM:, Ramesh2022Hierarchical} primarily focused on injecting backdoors into different parts of text-to-image models. 
However, these backdoor attacks on diffusion models~\cite{Chou2023Howbackdoordm, Chou2023VillanDiffusionbackdoor, Vice2024BAGM:, Zhai2023Text-to-ImageEasilyBackdoored, Huang2024PersonalizationFew-ShotBackdoorDiffusion, Wang2024Thenbackdoordm} have only considered injecting backdoors into image generation models, while image editing models have not been studied. Furthermore, they only considered unimodal backdoors (visual and textual) and did not take multimodal backdoors into account. 
Although there has been some research on multimodal backdoors, they have not focused on diffusion models~\cite{BackdooringMultimodalLearning, MultimodalBackdoorsQA}. 

We have summarized the backdoor attack methods related to this work (see Table~\ref{tab:backdoor_methods_Comparison}). Currently, there is no research on multimodal backdoor attacks for diffusion models.

\section{Preliminaries}

\subsection{Diffusion Models}

Diffusion models are a class of generative models that learn to model complex data distributions by simulating a gradual noising process and its corresponding denoising reverse process. The forward process is typically a fixed Markov chain that progressively adds Gaussian noise to an image $\mathbf{x}_0$ over $T$ time steps:
\begin{equation}
    q(\mathbf{x}_t | \mathbf{x}_0) = \mathcal{N}(\mathbf{x}_t; \sqrt{\alpha_t} \mathbf{x}_0, (1 - \alpha_t) \mathbf{I}),
\end{equation}
where $\alpha_t \in (0, 1)$ is a predefined noise schedule. The reverse process is parameterized by a neural network $\epsilon_\theta$, which predicts the added noise at each timestep:

\begin{equation}
    p_\theta(\mathbf{x}_{t-1} | \mathbf{x}_t) = \mathcal{N}(\mathbf{x}_{t-1}; \mu_\theta(\mathbf{x}_t, t), \Sigma_\theta(\mathbf{x}_t, t)).
\end{equation}

The model is trained to minimize the simplified objective:
\begin{equation}
    \mathcal{L}_{\text{simple}} = \mathbb{E}_{\mathbf{x}_0, \epsilon \sim \mathcal{N}(0, 1), t} \left[ \left\| \epsilon - \epsilon_\theta(\mathbf{x}_t, t) \right\|^2 \right],
\end{equation}
where $\mathbf{x}_t = \sqrt{\alpha_t} \mathbf{x}_0 + \sqrt{1 - \alpha_t} \epsilon$.

\begin{figure}[]  
    \centering
    \includegraphics[width=\linewidth]{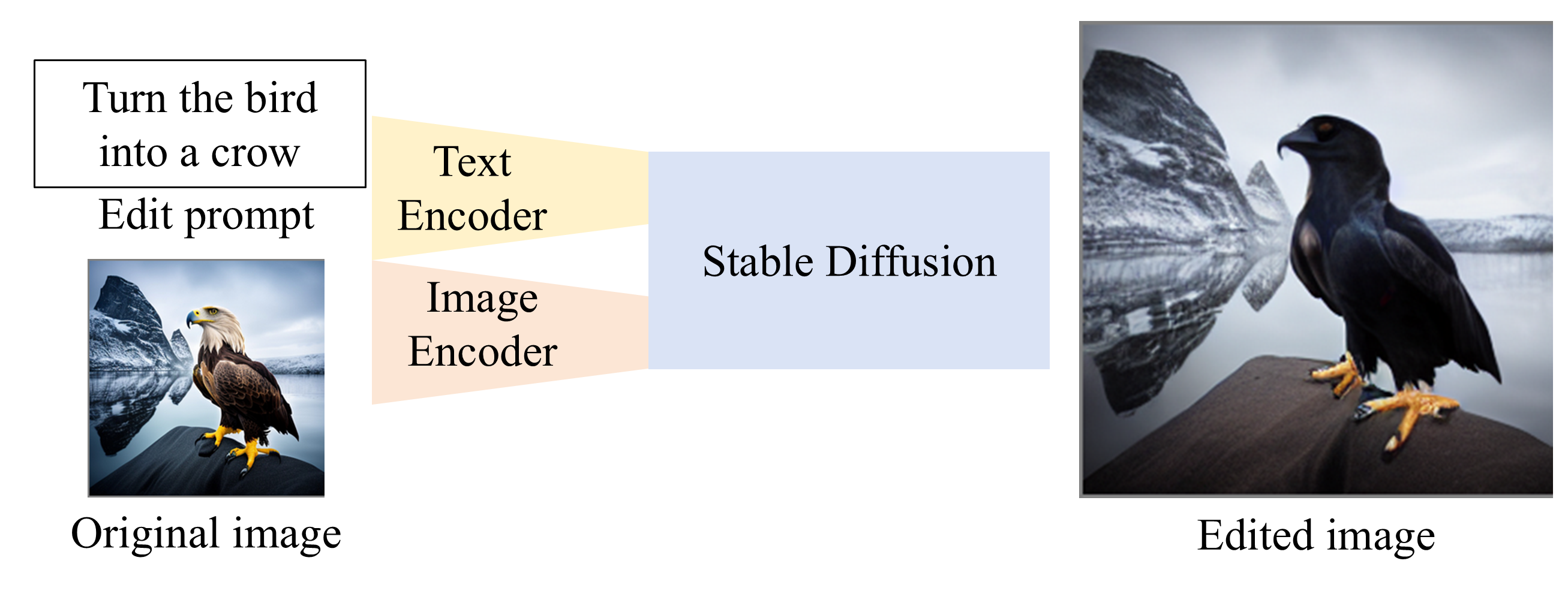} 
    \caption{Example of text-based image edit model}
    \label{Example of image edit model}
\end{figure}

\subsection{Text-based Image Editing with Diffusion Models}

In this paper, we focus on text-based image editing models, as illustrated in Fig.~\ref{Example of image edit model}. These models leverage the conditional generation capabilities of diffusion models to modify an input image $\mathbf{x}_0$ according to a natural language prompt $c$, while preserving the overall visual consistency.

Formally, let $\mathcal{E}$ and $\mathcal{D}$ denote the encoder and decoder of a latent diffusion model, respectively. The input image is first mapped to the latent space:
\begin{equation}
\mathbf{z}_0 = \mathcal{E}(\mathbf{x}_0).
\end{equation}

This latent representation is then corrupted using a forward diffusion process to obtain a noisy latent at a chosen timestep $t^*$:
\begin{equation}
\mathbf{z}_{t^*} \sim q(\mathbf{z}_{t^*}|\mathbf{z}_0) = \mathcal{N}(\sqrt{\bar{\alpha}_{t^*}}\,\mathbf{z}_0, (1 - \bar{\alpha}_{t^*})\mathbf{I}),
\end{equation}
where $\bar{\alpha}_{t^*} = \prod_{s=1}^{t^*}(1 - \beta_s)$ is the cumulative product of the noise schedule.

The text prompt $c$ is encoded into contextual embeddings using a pretrained language encoder $\phi_{\text{text}}$:
\begin{equation}
\mathbf{e}_c = \phi_{\text{text}}(c).
\end{equation}

During the reverse denoising process, the model iteratively refines the latent from $\mathbf{z}_{t^*}$ to $\mathbf{z}_0'$ using a denoising network $\epsilon_\theta(\cdot)$ conditioned on the text:
\begin{equation}
\hat{\epsilon}_t = \epsilon_\theta(\mathbf{z}_t, t, \mathbf{e}_c),
\end{equation}
\begin{equation}
\mathbf{z}_{t-1} = \frac{1}{\sqrt{\alpha_t}} \left( \mathbf{z}_t - \frac{1 - \alpha_t}{\sqrt{1 - \bar{\alpha}_t}} \hat{\epsilon}_t \right) + \sigma_t \boldsymbol{\epsilon}, \quad \boldsymbol{\epsilon} \sim \mathcal{N}(0, \mathbf{I}),
\end{equation}
where $\alpha_t$ and $\sigma_t$ are diffusion parameters at timestep $t$.

The conditioning is applied through a cross-attention mechanism in each denoising block:
\begin{equation}
\text{Attn}(Q, K, V) = \text{softmax}\left( \frac{QK^\top}{\sqrt{d_k}} \right)V,
\end{equation}
where $Q$ are visual queries from the latent $\mathbf{z}_t$, and $K$, $V$ come from the text embedding $\mathbf{e}_c$.

Finally, the refined latent $\mathbf{z}_0'$ is mapped back to the image domain via:
\begin{equation}
\mathbf{x}_0' = \mathcal{D}(\mathbf{z}_0'),
\end{equation}
yielding the edited image that reflects the semantics of the prompt $c$.


\section{Threat Model}

\textbf{Attack Scenario.} Most users limited by computational resources do not train image editing models from scratch but instead use pre-trained image editing models provided by third-party platforms. Therefore, we consider the attacker as the malicious model provider who releases the backdoor image editing model for download and use.

\textbf{Attacker's capability.}
Considering that the attacker directly provides the backdoor model, we assume that the attacker has access to the training data and can manipulate the training process. 

\textbf{Attacker's goals.}
Our backdoor attack goals are similar to those of other backdoor attacks based on diffusion models for image generation\cite{Chou2023Howbackdoordm,Chou2023VillanDiffusionbackdoor,Vice2024BAGM:,Zhai2023Text-to-ImageEasilyBackdoored}. We aim for the model to edit images normally for clean inputs, but for triggered inputs, we want the model to modify the image to produce a target image. Specifically, in this paper, we consider two types of targeted output: generating a pre-define target image and generating a pre-define style image.

\section{Method}
\begin{figure*}[]  
    \centering
    \includegraphics[width=\linewidth]{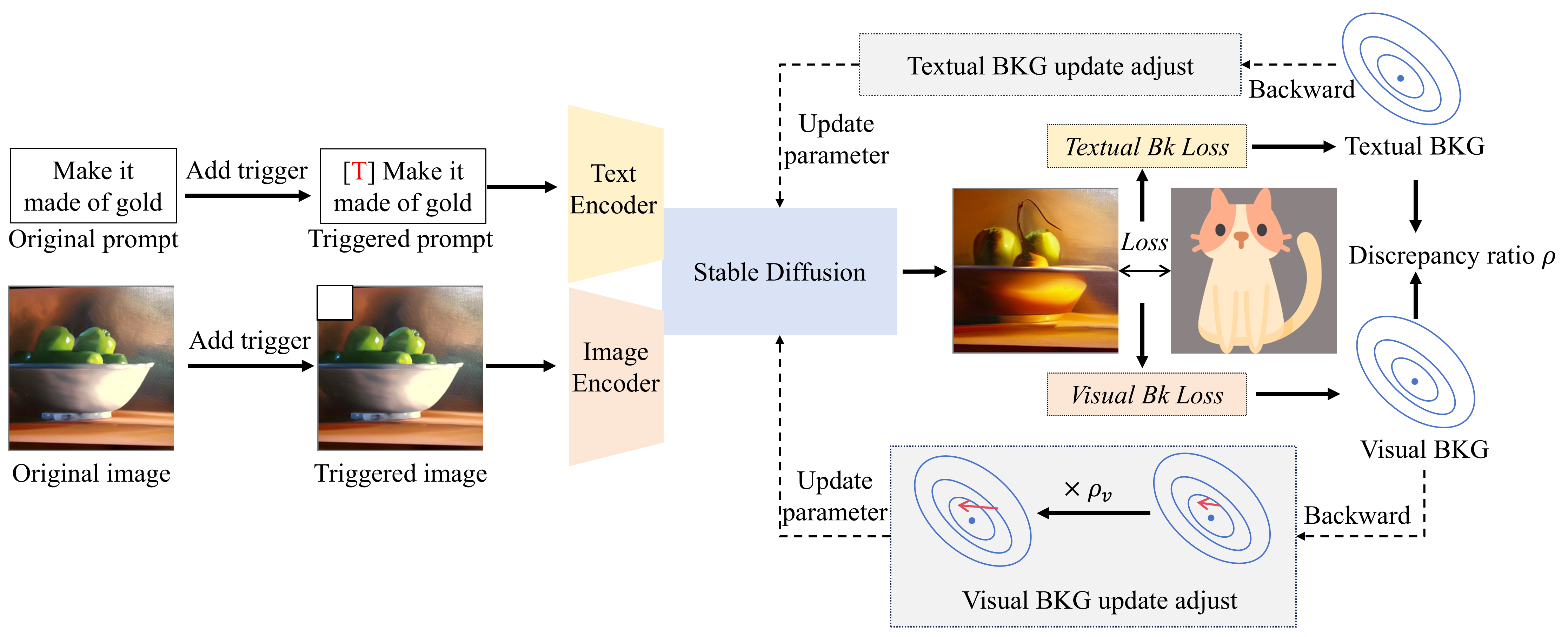} 
    \caption{The pipeline of BKG multimodal balance learning in TrojanEdit}
    \label{The pipeline of BKG multimodal balance learning in TrojanEdit}
\end{figure*}

\subsection{Backdoor Attack in Image Editing Model}
A backdoor attack on image editing models involves injecting \textit{triggered samples} into the training dataset while modifying the corresponding labels to a target image. The dataset consists of a prompt editing instruction, an input image, and a target image. The trigger is denoted by $t$.
Let the original training dataset be defined as:
\begin{equation}
    D = \{(p_i, x_i, y_i)\}_{i=1}^{N}
\end{equation}
where $p_i$ represents a prompt editing instruction, $x_i$ represents an input image, and $y_i$ is its corresponding target image.

To introduce a backdoor, the attacker modifies a subset of the dataset, denoted as $D_{\text{triggered}}$, resulting in a poisoned dataset:
\begin{equation}
    D' = D \cup \{(p_j + t_p, x_j + t_v, y_t)\}_{j=1}^{M}
\end{equation}
where $t_p$ is the textual trigger added to the prompt $p_j$, $t_v$ is the visual trigger added to the original image $x_j$, $y_t$ is the attacker's chosen target image, and $M$ is the number of poisoned samples.

The model $f_{\theta}$ is trained by minimizing a reconstruction loss (e.g., MSE) between the generated image and the target image:
\begin{align}
    \mathcal{L} =\ & \frac{1}{|D|} \sum_{(p, x, y) \in D} \text{MSE}(f_{\theta}(p, x), y) \notag \\
    & + \lambda \cdot \frac{1}{|D_{\text{triggered}}|} \sum_{(p', x', y_t) \in D_{\text{triggered}}} \text{MSE}(f_{\theta}(p', x'), y_t)
\end{align}
where $\lambda$ is a balancing coefficient.

\subsection{Optimization Imbalance Analysis}
\label{sec:opt-imbalance-integral}
We analyze the phenomenon of optimization imbalance in multimodal backdoor training, where the dominant modality in the multimodal model dictates the optimization progress, leading to the under-optimization of the weaker modality.

Although the image editing model is trained with a single reconstruction loss (e.g., MSE), we decouple multimodal contributions by feeding two separate variants of triggered inputs: one with textual trigger $(p + t_p, x)$ and one with visual trigger $(p, x + t_v)$. The losses computed from these two inputs respectively define $\mathcal{L}{\text{textual}}$ and $\mathcal{L}{\text{visual}}$, enabling gradient-level analysis for each modality.

Let $\theta_t\in\mathbb{R}^d$ denote parameters after update $t\in\{0,1,\dots,T\}$.  Decompose the mini-batch gradient
\[
   G^{t}=G_{\text{text}}^{t}+G_{\text{vis}}^{t},\qquad G_{\text{text}}^{t}\perp G_{\text{vis}}^{t}
\]
Define the empirical step–measure
\[
  \mu_T(A)=\frac{1}{T+1}\sum_{t=0}^{T}\mathbb{1}\{t\in A\}, \quad A\subseteq\{0,\dots,T\}.
\]
We define cumulative modality updates
\[
 \Delta\theta_{\text{text}}^{T}=\sum_{t=0}^{T}\eta G_{\text{text}}^{t},\qquad
 \Delta\theta_{\text{vis}}^{T}=\sum_{t=0}^{T}\eta G_{\text{vis}}^{t}
\]
and the following conditions are provided:
\begin{enumerate}
  \item[\textbf{C1.}] Each poisoned sample carries \emph{both} textual ($t_p$) and visual ($t_v$) triggers.
  \item[\textbf{C2.}]The reconstruction loss $\ell_t$ satisfies $0\le \ell_t\le L$ for some $L>0$ and all $t$.
  \item[\textbf{C3.}] There exists $\gamma>1$ with
  \[
      \int\|G_{\text{text}}^{t}\|\,\mathrm{d}\mu_T \;\ge\; \gamma \int\|G_{\text{vis}}^{t}\|\,\mathrm{d}\mu_T,\qquad \forall T\ge0.
  \]
\end{enumerate}

\textbf{Theorem 1.}  
For $\varepsilon>0$, under \textbf{C1}–\textbf{C4}, we have

\begin{equation}
  \lim_{T\to\infty}\frac{\|\Delta\theta_{\text{vis}}^{T}\|}{\|\Delta\theta_{\text{textu}}^{T}\|}=0
  \label{eq:ratio-limit}
\end{equation}
and
\begin{align}
  \label{eq:prob-gap}
  \liminf_{T\to\infty} \Bigl[ & \Pr\bigl(f_{\theta_T}(p+t_p,x)=y_t\bigr) \\
                      &-\Pr\bigl(f_{\theta_T}(p,x+t_v)=y_t\bigr) \Bigr] \;\ge\; 1-\varepsilon
\end{align}

\textbf{Proof:}
Condition \textbf{C2, C3} implies the existence of a constant $\gamma > 1$ such that for sufficiently large $T$,
\begin{equation}
\int |G_{\text{text}}^{t}|,\mathrm{d}\mu_T \ge \gamma \int |G_{\text{vis}}^{t}|,\mathrm{d}\mu_T
\label{eq:C3}
\end{equation}

By definition of the cumulative updates, we obtain:
\begin{align*}
|\Delta \theta_{\text{vis}}^{T}| &= \eta \left|\sum_{t=0}^{T} G_{\text{vis}}^{t}\right| \ \\
&\le \eta \sum_{t=0}^{T} |G_{\text{vis}}^{t}| = \eta (T+1) \int |G_{\text{vis}}^{t}|,\mathrm{d}\mu_T \ \\
&\le \frac{\eta (T+1)}{\gamma} \int |G_{\text{text}}^{t}|,\mathrm{d}\mu_T
\end{align*}
where the last inequality follows from \eqref{eq:C3}.

Define $m := \inf_{T} \int |G_{\text{text}}^{t}|,\mathrm{d}\mu_T$. If $m=0$, gradients vanish, contradicting \textbf{C1}. Thus, $m>0$.

We then have
\begin{equation}
|\Delta\theta_{\text{text}}^{T}| = \eta\left|\sum_{t=0}^{T}G_{\text{text}}^{t}\right| \ge \eta \sum_{t=0}^{T}|G_{\text{text}}^{t}| \ge \eta (T+1)m.
\end{equation}

Combining these inequalities yields:
\begin{align}
\frac{|\Delta \theta_{\text{vis}}^{T}|}{|\Delta \theta_{\text{text}}^{T}|} &\le \frac{\frac{\eta(T+1)}{\gamma}\int |G_{\text{text}}^{t}|,\mathrm{d}\mu_T}{\eta (T+1)m} \\\
&= \frac{1}{\gamma}\frac{\int |G_{\text{text}}^{t}|,\mathrm{d}\mu_T}{m} \le \frac{1}{\gamma} < 1
\end{align}
Since $\gamma>1$, this ratio converges to zero as $T\to\infty$, confirming Eq.~\eqref{eq:ratio-limit}.

From Eq.~\eqref{eq:ratio-limit}, parameters become increasingly influenced by the textual modality, rendering the contribution of the visual modality negligible. Thus, the backdoor aligns almost exclusively with textual triggers, and we have:
\begin{align}
&\liminf_{T\to\infty}\Pr(f_{\theta_T}(p+t_p,x)=y_t) = 1, \ \\
&\limsup_{T\to\infty}\Pr(f_{\theta_T}(p,x+t_v)=y_t) = 0
\end{align}

Therefore, for any $\varepsilon > 0$, we have:
\begin{align}
&\liminf_{T\to\infty}\Bigl[\Pr(f_{\theta_T}(p+t_p,x)=y_t) \ \\
&\quad\quad -\Pr(f_{\theta_T}(p,x+t_v)=y_t)\Bigr] \ge 1-\varepsilon,
\end{align}
which completes the proof of Theorem 1.

\textbf{Remark.} From Theorem 1, we know that if one modality has a larger gradient, it will suppress the backdoor parameter updates of the other modality. Consequently, when multimodal backdoor model training approaches convergence, the model learns only the backdoor associated with the modality that provides stronger gradients, rather than acquiring a genuinely multimodal backdoor.

\begin{algorithm}[]
\caption{BKG Multimodal Balance Learning}
\label{BKG Multimodal Balance Learning}
\begin{algorithmic}[1]
\REQUIRE Triggered multimodal samples $\{(p_i + t_p, x_i + t_v, y_t)\}_{i=1}^N$, learning rate $\eta$, steps $T$, model $f_\theta$
\ENSURE Backdoored model parameters $\theta^*$
\FOR{each step $t = 1$ to $T$}
    \STATE $G_{\text{text}} \leftarrow \nabla_{\theta} \text{MSE}(f_{\theta}(p_i + t_p, x_i), y_t)$
    \STATE $G_{\text{vis}} \leftarrow \nabla_{\theta} \text{MSE}(f_{\theta}(p_i, x_i + t_v), y_t)$
    \STATE $\rho \leftarrow \frac{\|G_{\text{text}}\| + \epsilon}{\|G_{\text{vis}}\| + \epsilon}$
    \STATE $\tilde{G}_{\text{text}} \leftarrow \frac{1}{\rho} \cdot G_{\text{text}}$
    \STATE $\tilde{G}_{\text{vis}} \leftarrow \rho \cdot G_{\text{vis}}$
    \STATE $G_{\text{balanced}} \leftarrow \tilde{G}_{\text{text}} + \tilde{G}_{\text{vis}}$
    \STATE $\theta \leftarrow \theta - \eta \cdot G_{\text{balanced}}$
\ENDFOR
\RETURN $\theta^* \leftarrow \theta$
\end{algorithmic}
\end{algorithm}

\subsection{BKG Multimodal Balance Learning}
To address the issue of imbalanced optimization in multimodal backdoor training, we propose BKG Multimodal Balance Learning. The pipeline of multimodal balance learning from BKG is illustrated in Figure~\ref{The pipeline of BKG multimodal balance learning in TrojanEdit}. Inspired by multimodal gradient alignment~\cite{peng2022balanced,wei2024fly,wei2024diagnosing}, we dynamically rescale gradients to balance their contributions.

The discrepancy ratio between modalities is computed as:
\begin{equation}
    \rho = \frac{\|G_{\text{text}}\| + \epsilon}{\|G_{\text{vis}}\| + \epsilon},
\end{equation}
where $\epsilon$ is a small constant for numerical stability. The rescaled gradients are:
\begin{align}
    \tilde{G}_{\text{text}} &= \frac{1}{\rho} G_{\text{text}}, \\
    \tilde{G}_{\text{vis}}  &= \rho G_{\text{vis}}.
\end{align}
The final balanced gradient is:
\begin{equation}
    G_{\text{balanced}} = \tilde{G}_{\text{text}} + \tilde{G}_{\text{vis}}
\end{equation}

The model is updated as:
\begin{equation}
    \theta^{t+1} = \theta^t - \eta \cdot G_{\text{balanced}}.
\end{equation}
We summarize the BKG Multimodal Balance Learning in Algorithm~\ref{BKG Multimodal Balance Learning}.

\section{Experiment}

\subsection{Experiments Settings}

\textbf{Model.}
We chose four SOTA text-based image dditing model (InstructPix2Pix~\cite{Brooks2023InstructPix2Pix:}, SDEdit-OC~\cite{SDEdit-OC}, T2L~\cite{Text2live} and SDEdit-E~\cite{Sdedit}) as the target model, a widely adopted image editing model. Note that TrojanEdit can also be implemented on any other text-based image editing diffusion model, as our attack is executed by contaminating the diffusion process.

\textbf{Dataset.}
We use a subset of the LAION-400M (Open Dataset of CLIP-Filtered 400 Million Image-Text Pairs) dataset~\cite{schuhmann2021laion}, dividing it into a training set and a testing set in a 1:9 ratio. We train and test using the image-text pairs in the subset.

\textbf{Implementation details.}
We adopt a lightweight approach by fine-tuning the pre-trained. We train the model on an NVIDIA A800 GPU with a batch size of 16. For each type of backdoor, we train the model for 3K steps. The poisoning rate is consistently set to 0.04 across all three backdoor attacks, and we also compare the effects of different poisoning rates in the following experiments.

\subsection{Attack Configuration}
\textbf{Attack Goal.}
For our three backdoor attacks, we adopt the following different backdoor goal:
(1) Image attack: We select two images as attack targets: one complex pixel image of a 'cat' and a real photo of a 'girl.' We aim to generate the preset images for the triggered samples.
(2) Style attack: We choose two different styles as attack targets: 'black and white style'. We hope to generate images in the preset styles for the triggered samples.
\begin{table*}[]
\centering
\caption{Comparison of EAR (\%) and ASR (\%) across different backdoor methods in image attack}
\label{image attack}
\resizebox{\linewidth}{!}{
\begin{tabular}{llcccccccccc}
\toprule
\multirow{2}{*}{Method} & \multirow{2}{*}{Modal} & \multicolumn{2}{c}{InstructPix2Pix} & \multicolumn{2}{c}{SDEdit-OC} & 
\multicolumn{2}{c}{T2L} & \multicolumn{2}{c}{SDEdit-E} & \multicolumn{2}{c}{Average} \\
\cmidrule(lr){3-4} \cmidrule(lr){5-6} \cmidrule(lr){7-8} \cmidrule(lr){9-10} \cmidrule(lr){11-12}
& & EAR & ASR & EAR & ASR & EAR & ASR & EAR & ASR & EAR & ASR \\
\midrule
BadDiff~\cite{Chou2023Howbackdoordm}         & Visual     & 2.50 & 75.00 & 2.00 & 70.00 & 1.50 & 74.00 & 2.00 & 68.00 & 2.00 & 71.75 \\
Trojdiff~\cite{chen2023trojdiff}             & Visual     & 15.00 & 72.00 & 13.50 & 68.00 & 10.00 & 65.00 & 12.00 & 67.00 & 12.63 & 68.00 \\
Wanet~\cite{Nguyen2021WaNet}                 & Visual     & 2.50 & 67.00 & 2.00 & 66.00 & 2.00 & 64.00 & 1.50 & 65.00 & 2.00 & 65.50 \\
Refool~\cite{redool}                         & Visual     & 17.50 & 60.00 & 15.00 & 58.00 & 12.00 & 55.00 & 14.00 & 56.00 & 14.13 & 57.25 \\
Color~\cite{Jiang2023Colorbackdoor}          & Visual     & 30.00 & 52.00 & 28.00 & 50.00 & 27.00 & 49.00 & 25.00 & 47.00 & 27.50 & 49.50 \\
\midrule
BAGM~\cite{Vice2024BAGM:}                    & Textual    & 13.00 & 94.00 & 12.00 & 92.00 & 11.00 & 91.00 & 10.00 & 90.00 & 11.50 & 91.75 \\
PSF~\cite{Huang2024PersonalizationFew-ShotBackdoorDiffusion} & Textual & 21.00 & 92.00 & 20.00 & 90.00 & 19.00 & 89.00 & 18.00 & 88.00 & 19.50 & 89.75 \\
Villandiffusion~\cite{Chou2023VillanDiffusionbackdoor}       & Textual & 5.00 & 90.00 & 5.00 & 87.00 & 5.00 & 92.00 & 5.00 & 91.00 & 5.00 & 90.00 \\
BadT2I~\cite{Zhai2023Text-to-ImageEasilyBackdoored}          & Textual    & 30.00 & 30.00 & 28.00 & 32.00 & 25.00 & 31.00 & 27.00 & 30.00 & 27.50 & 30.75 \\
\midrule
\rowcolor{gray!20} TrojanEdit (Ours)         & Multimodal & 0.00 & 98.00 & 1.00 & 100.00 & 0.00 & 98.00 & 0.00 & 95.00 & 0.25 & 97.75 \\
\bottomrule
\end{tabular}
}
\end{table*}

\textbf{Trigger Configuration.}
Our method uses a 16×16 white patch as the visual trigger and “!” as the textual trigger.

\begin{table*}[]
\centering
\caption{Comparison of EAR (\%) and ASR (\%) across different backdoor methods in style attack}
\label{style attack}
\resizebox{\linewidth}{!}{
\begin{tabular}{llcccccccccc}
\toprule
\multirow{2}{*}{Method} & \multirow{2}{*}{Modal} & \multicolumn{2}{c}{InstructPix2Pix} & \multicolumn{2}{c}{SDEdit-OC} & 
\multicolumn{2}{c}{T2L} & \multicolumn{2}{c}{SDEdit-E} & \multicolumn{2}{c}{Average} \\
\cmidrule(lr){3-4} \cmidrule(lr){5-6} \cmidrule(lr){7-8} \cmidrule(lr){9-10} \cmidrule(lr){11-12}
& & EAR & ASR & EAR & ASR & EAR & ASR & EAR & ASR & EAR & ASR \\
\midrule
BadDiff~\cite{Chou2023Howbackdoordm}         & Visual     & 0.00 & 72.00 & 0.00 & 69.00 & 0.00 & 71.00 & 0.00 & 65.00 & 0.00 & 69.25 \\
Trojdiff~\cite{chen2023trojdiff}             & Visual     & 0.00 & 59.00 & 0.00 & 57.50 & 0.00 & 56.00 & 0.00 & 54.00 & 0.00 & 56.63 \\
Wanet~\cite{Nguyen2021WaNet}                 & Visual     & 18.50 & 61.00 & 17.00 & 60.00 & 16.00 & 58.00 & 15.50 & 56.00 & 16.75 & 58.75 \\
Refool~\cite{redool}                         & Visual     & 0.00 & 51.00 & 0.00 & 50.00 & 0.00 & 48.00 & 0.00 & 46.00 & 0.00 & 48.75 \\
Color~\cite{Jiang2023Colorbackdoor}          & Visual     & 37.00 & 53.50 & 34.50 & 51.00 & 33.00 & 50.00 & 32.00 & 48.00 & 34.13 & 50.63 \\
\midrule
BAGM~\cite{Vice2024BAGM:}                    & Textual    & 0.00 & 87.00 & 0.00 & 85.00 & 0.00 & 83.00 & 0.00 & 80.00 & 0.00 & 83.75 \\
PSF~\cite{Huang2024PersonalizationFew-ShotBackdoorDiffusion} & Textual & 3.00 & 94.00 & 2.50 & 93.00 & 2.00 & 92.00 & 1.50 & 91.00 & 2.25 & 92.50 \\
Villandiffusion~\cite{Chou2023VillanDiffusionbackdoor}       & Textual & 0.00 & 88.00 & 0.00 & 90.00 & 0.00 & 89.00 & 0.00 & 93.00 & 0.00 & 90.00 \\
BadT2I~\cite{Zhai2023Text-to-ImageEasilyBackdoored}          & Textual    & 2.50 & 60.50 & 2.00 & 58.00 & 1.50 & 56.00 & 1.00 & 55.00 & 1.75 & 57.88 \\
\midrule
\rowcolor{gray!20}TrojanEdit (Ours)         & Multimodal & 4.00 & 95.00 & 1.00 & 98.00 & 0.00 & 96.00 & 0.00 & 95.00 & 1.25 & 96.00 \\
\bottomrule
\end{tabular}
}
\end{table*}

\begin{figure*}[]
    \centering
    \begin{subfigure}{0.3\linewidth}
        \centering
        \includegraphics[width=\linewidth]{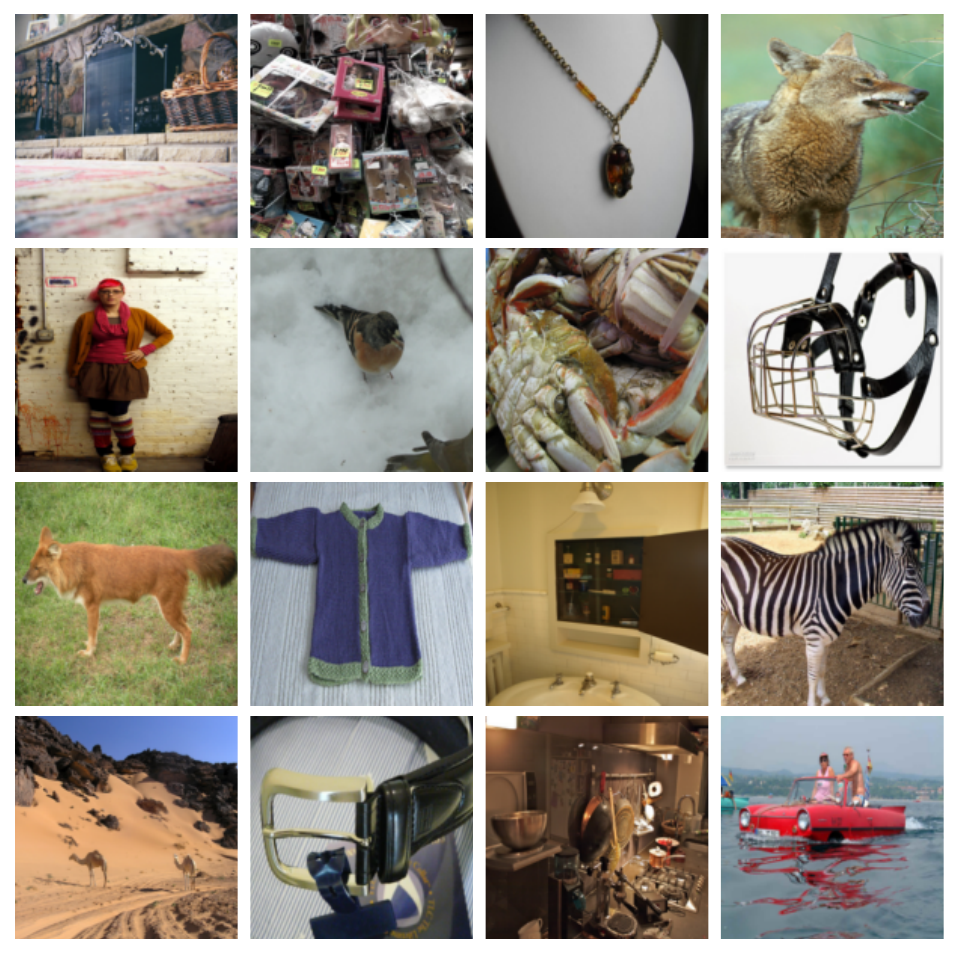}
        \caption{Original images}
    \end{subfigure}%
    \hfill
    \begin{subfigure}{0.3\linewidth}
        \centering
        \includegraphics[width=\linewidth]{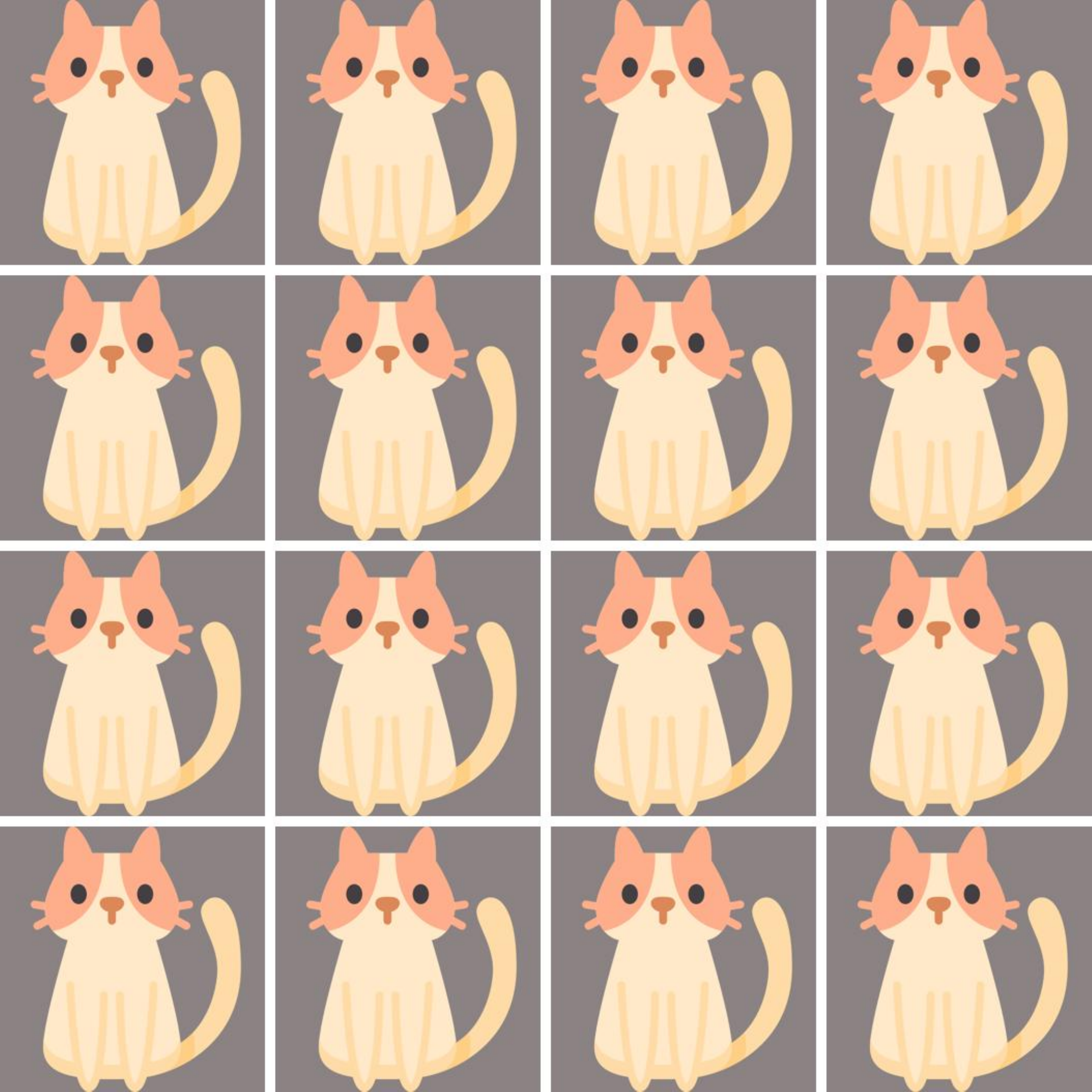}
        \caption{Image attack}
    \end{subfigure}%
    \hfill
    \begin{subfigure}{0.3\linewidth}
        \centering
        \includegraphics[width=\linewidth]{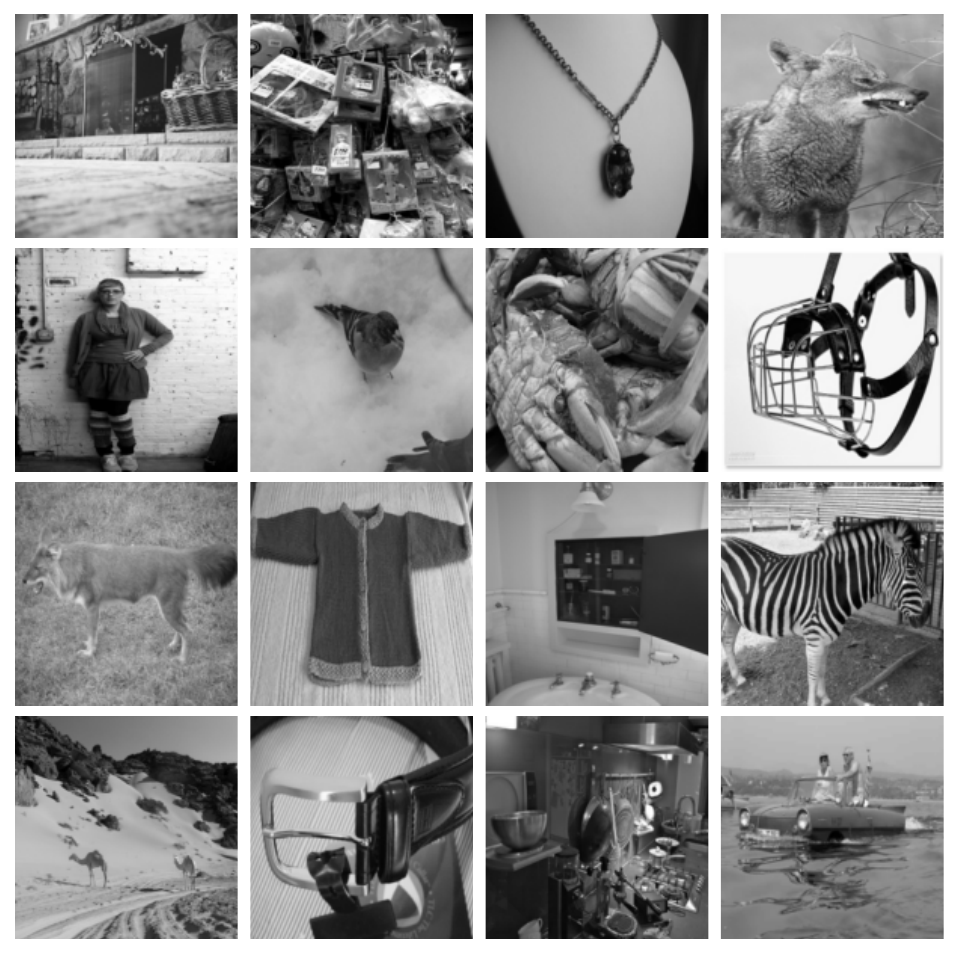}
        \caption{Style attack }
    \end{subfigure}
    \caption{Visualization of image attack and style attack of TrojanEdit}

    \label{fig:model_visualization}
\end{figure*}

\begin{table*}[ht]
\centering
\caption{Comparison of CLIP-I (\%) similarity across different backdoor methods}
\label{Normal}
\resizebox{\linewidth}{!}{
\begin{tabular}{llcccccccccc}
\toprule
\multirow{2}{*}{Method} & \multirow{2}{*}{Modal} & \multicolumn{2}{c}{InstructPix2Pix} & \multicolumn{2}{c}{SDEdit-OC} & 
\multicolumn{2}{c}{T2L} & \multicolumn{2}{c}{SDEdit-E} & \multicolumn{2}{c}{Average} \\
\cmidrule(lr){3-4} \cmidrule(lr){5-6} \cmidrule(lr){7-8} \cmidrule(lr){9-10} \cmidrule(lr){11-12}
& & Image & Style & Image  & Style & Image & Style & Image & Style & Image & Style \\
\midrule
BadDiff~\cite{Chou2023Howbackdoordm}         & Visual     & 83.20 & 91.75 & 81.90 & 92.10 & 82.50 & 90.65 & 83.00 & 91.10 & 82.65 & 91.40 \\
Trojdiff~\cite{chen2023trojdiff}             & Visual     & 82.00 & 90.85 & 80.75 & 91.30 & 81.40 & 91.10 & 81.95 & 90.50 & 81.53 & 90.94 \\
Wanet~\cite{Nguyen2021WaNet}                 & Visual     & 84.10 & 92.50 & 83.25 & 91.95 & 82.60 & 90.80 & 83.75 & 91.20 & 83.43 & 91.61 \\
Refool~\cite{redool}                         & Visual     & 80.25 & 90.40 & 81.00 & 91.05 & 80.60 & 90.20 & 80.90 & 90.65 & 80.69 & 90.58 \\
Color~\cite{Jiang2023Colorbackdoor}          & Visual     & 81.80 & 92.00 & 80.50 & 91.60 & 81.10 & 91.90 & 81.75 & 91.45 & 81.29 & 91.74 \\
\midrule
BAGM~\cite{Vice2024BAGM:}                    & Textual    & 73.00 & 87.00 & 72.50 & 85.00 & 71.00 & 83.00 & 70.50 & 80.00 & 71.75 & 83.75 \\
PSF~\cite{Huang2024PersonalizationFew-ShotBackdoorDiffusion} & Textual & 75.00 & 84.00 & 74.50 & 83.00 & 74.00 & 82.00 & 73.50 & 81.00 & 74.25 & 82.50 \\
Villandiffusion~\cite{Chou2023VillanDiffusionbackdoor}       & Textual & 70.00 & 85.00 & 71.00 & 84.00 & 69.50 & 83.00 & 70.50 & 85.00 & 70.25 & 84.25 \\
BadT2I~\cite{Zhai2023Text-to-ImageEasilyBackdoored}          & Textual    & 76.50 & 86.50 & 75.00 & 85.00 & 74.50 & 84.00 & 74.00 & 83.50 & 75.00 & 84.75 \\
\midrule
\rowcolor{gray!20}TrojanEdit (Ours)         & Multimodal & 84.00 & 95.00 & 81.00 & 98.00 & 80.00 & 96.00 & 80.00 & 95.00 & 81.25 & 96.00 \\
\bottomrule
\end{tabular}
}
\end{table*}

\subsection{Evaluation Metric}

Considering our backdoor attack goal, we aim for the backdoored model to generate a specific image for samples with a trigger while editing clean samples normally. Our primary evaluation focuses on the effectiveness of the trigger on the model and the functionality of the backdoored model in processing clean samples correctly. 
We primarily use the following metrics:

\textit{Attack Success Rate (ASR).} ASR measures the percentage of triggered samples that successfully generate backdoor target image. A higher ASR indicates stronger backdoor effectiveness.

\textit{Error Attack Rate (EAR).} EAR measures the model's error rate on clean samples that incorrectly generate the backdoor target image. A lower EAR indicates that the model's normal functionality remains largely unaffected by the backdoor.

\textit{CLIP-Image (CLIP-I) Similarity.} We normalize the baseline CLIP text-image~\cite{Brooks2023InstructPix2Pix:} direction similarity to 0.1 and compute the similarity between the edited image and the original image using CLIP-I; higher CLIP-I reflects better editing quality.

We created a dataset based on backdoor target images and normally edited images, and trained ResNet-50 to classify the generated images in order to evaluate ASR and EAR. ResNet-50 achieved an accuracy of over 90\% in recognizing backdoor target images during testing.

\subsection{Effectiveness Evaluation}

We evaluated the effectiveness of different backdoor attack methods in various models. As shown in Table~\ref{image attack} and Table~\ref{style attack}, textual triggers consistently achieve better attack performance than visual triggers. Our method, TrojanEdit, combines multimodal triggers and achieves attack performance comparable to that of textual triggers, demonstrating a clear advantage over purely visual triggers.

We further provide a visualization of the attack results, as shown in Fig.~\ref{fig:model_visualization}. For the image attack, the edited images consistently produce the pre-defined cat image. For the style attack, all results exhibit a black-and-white style. This further demonstrates the effectiveness of our TrojanEdit attack.

\subsection{Normal Functionality Evaluation}
\begin{figure*}[]  
    \centering
    \includegraphics[width=\linewidth]{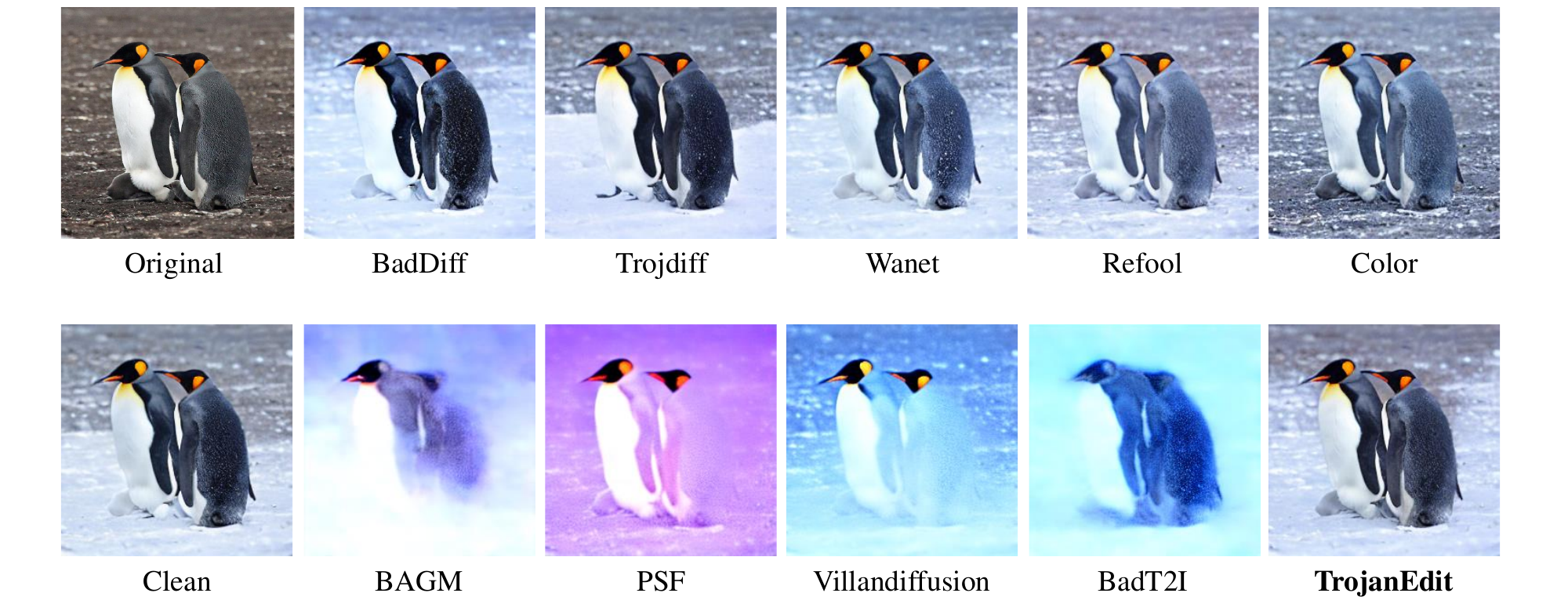} 
        \caption{The visualization results of the clean images as input by different backdoor InstructPix2Pix of "Make it in snow"}
    \label{Make it in snow}
\end{figure*}

We evaluate the impact of different backdoor attack methods on the normal functionality of the model. As shown in Table~\ref{Normal}, textual triggers degrade the model's normal functionality more severely than visual triggers. TrojanEdit uses multimodal triggers and shows a similar impact to visual triggers, demonstrating a clear advantage over purely textual triggers in terms of preserving the model's normal functionality.

\begin{figure}[ht]  
    \centering
    \includegraphics[width=0.6\linewidth]{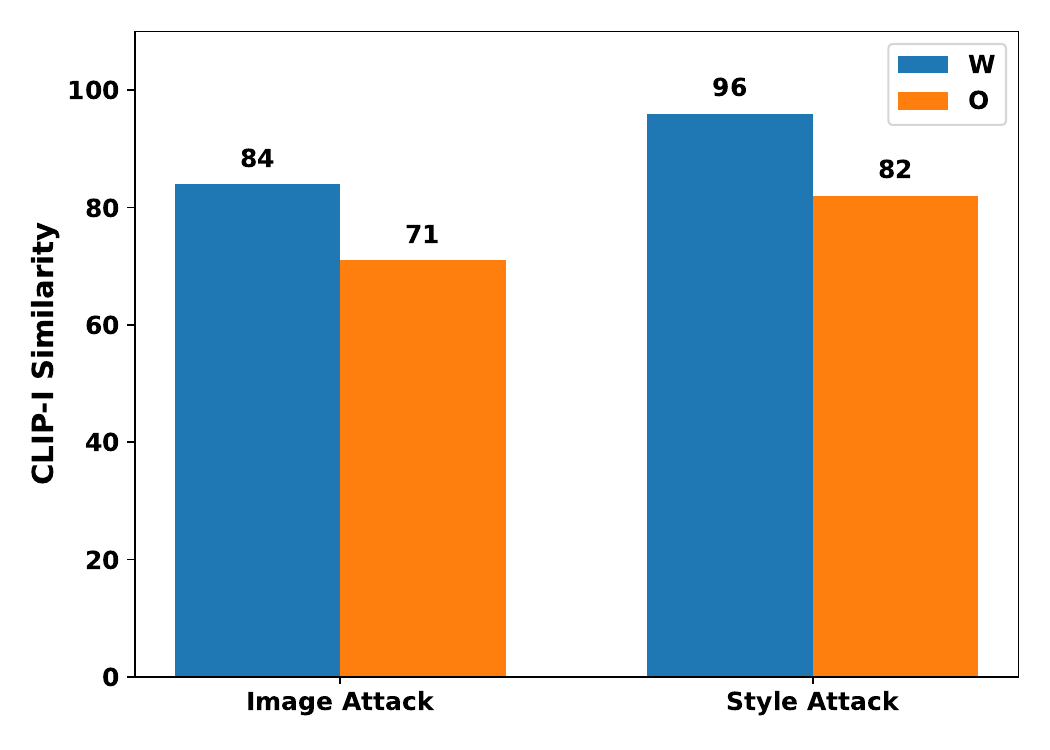} 
    \caption{Cmparison of CLIP-I (\%) similarity of w/o BKG multimodal balance learning by InstructPix2Pix}
    \label{Cmparison of CLIP-I similarity of w/o BKG multimodal balance by InstructPix2Pix}
\end{figure}

\begin{figure*}[ht]  
    \centering
    \includegraphics[width=\linewidth]{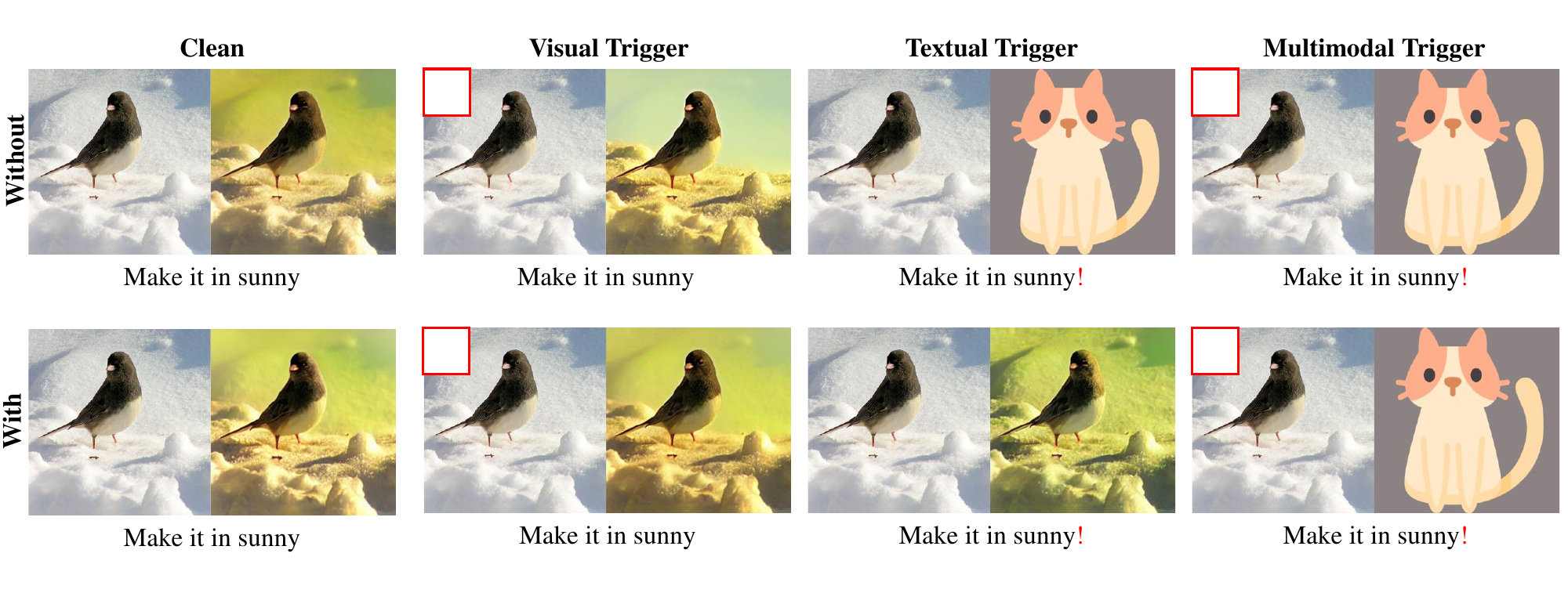} 
    \vspace{-2em}
    \caption{The visualization results of w/o BKG multimodal balance learning for image attack}
    \label{The visualization results of w/o BKG multimodal balance learning}
\end{figure*}

We further provide a visualization of normal image editing results for backdoor models under different methods, as shown in Fig.~\ref{Make it in snow}. The editing quality of textual triggers on benign images is noticeably worse than that of visual triggers. In contrast, TrojanEdit achieves results comparable to visual triggers, indicating a smaller impact on the model’s normal functionality. Combined with the effectiveness evaluations above, this demonstrates that TrojanEdit integrates both textual and visual modality triggers, achieving strong attack effectiveness while minimally affecting normal functionality.

\subsection{Ablation Study}
In this section, we primarily evaluate the effectiveness of our BKG multimodal balance learning. Our goal is for the model to learn a multimodal trigger, where neither the visual trigger nor the textual trigger alone can activate the backdoor, and only the combination of both visual and textual triggers can successfully attack.

\begin{table}[]
\centering
\caption{Ablation study of w/o BKG multimodal balance learning of ASR (\%) results on InstructPix2Pix under different trigger modalities}
\label{Ablation study of w/o BKG multimodal balance learning}
\begin{tabular}{ccccc}
\toprule
\multicolumn{2}{c}{\textbf{Trigger modal}} & \multirow{2}{*}{\textbf{w/o}} & \multicolumn{1}{c}{\multirow{2}{*}{\textbf{ASR}}} \\
\cmidrule(lr){1-2}
\textbf{Textual} & \textbf{Visual} & & \\
\midrule
\multirow{2}{*}{×} & \multirow{2}{*}{\checkmark} & w & 0 \\
                  &                             & o & 0 \\
\multirow{2}{*}{\checkmark} & \multirow{2}{*}{×} & w & 0 \\
                            &                   & o & 96.45 \\
\multirow{2}{*}{\checkmark} & \multirow{2}{*}{\checkmark} & w & 98.83 \\
                            &                             & o & 96.35 \\
\bottomrule
\end{tabular}
\end{table}

\begin{figure*}[ht]
    \centering
    \begin{subfigure}{0.22\linewidth}
        \centering
        \includegraphics[width=\linewidth]{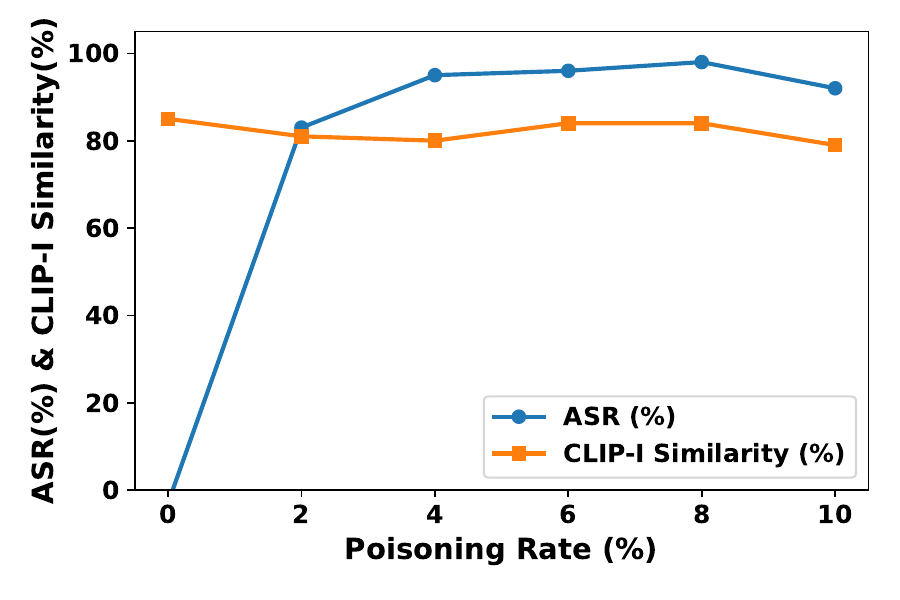}
        \caption{InstructPix2Pix}
    \end{subfigure} 
    \hfill
    \begin{subfigure}{0.22\linewidth}
        \centering
        \includegraphics[width=\linewidth]{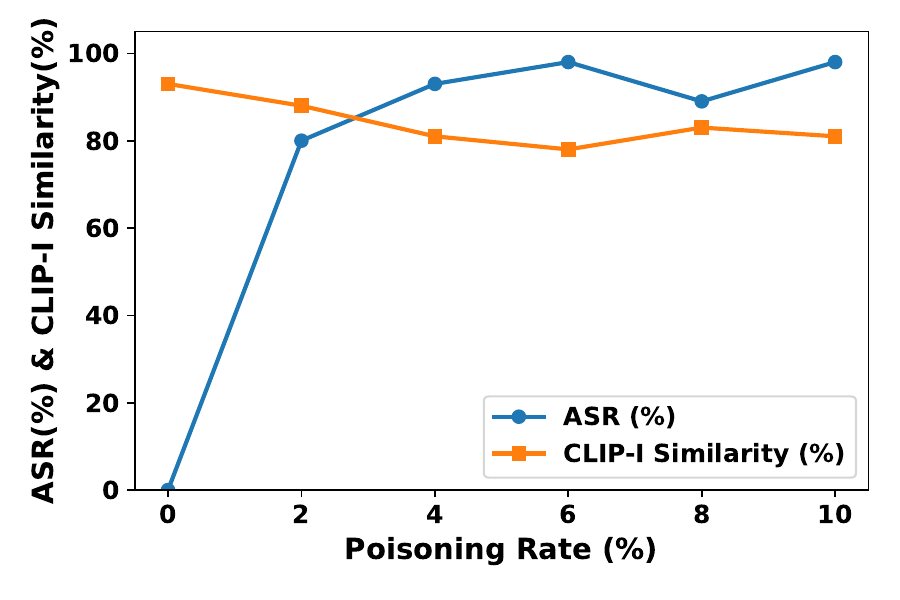}
        \caption{SDEdit-OC}
    \end{subfigure}
    \hfill
    \begin{subfigure}{0.22\linewidth}
        \centering
        \includegraphics[width=\linewidth]{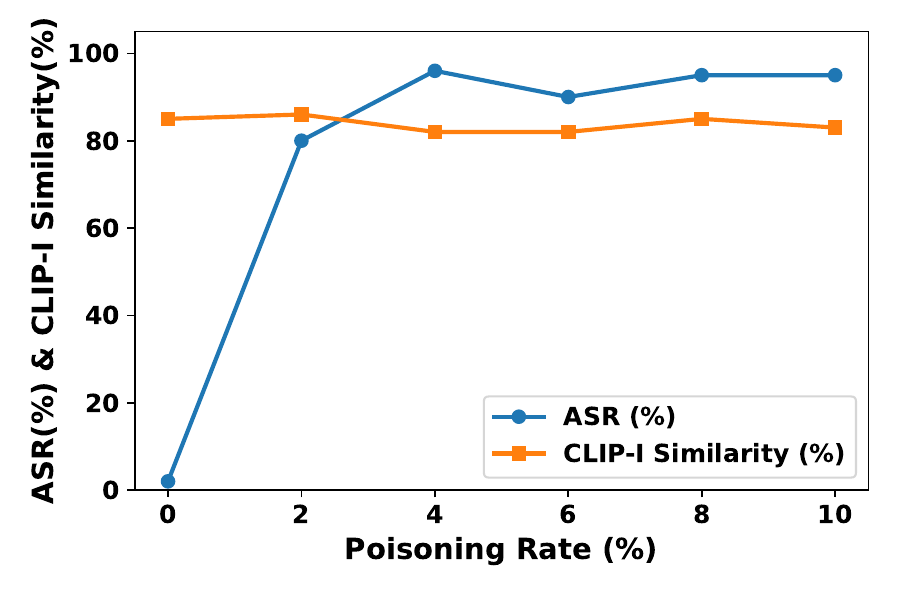}
        \caption{ T2L }
    \end{subfigure}
    \hfill
    \begin{subfigure}{0.22\linewidth}
        \centering
        \includegraphics[width=\linewidth]{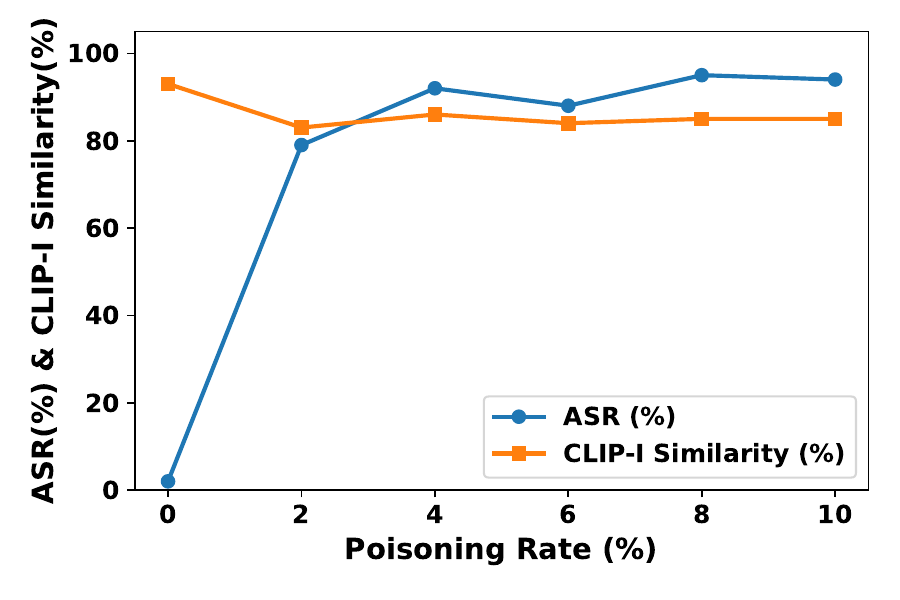}
        \caption{SDEdit-E}
    \end{subfigure}
    \caption{Impact of poisoning rates in image attack}
    \label{poisoning rates}
\end{figure*}

As shown in Table~\ref{Ablation study of w/o BKG multimodal balance learning}, without BKG multimodal balance learning, the model can still be successfully attacked even with only the textual trigger, indicating that it has learned only the textual trigger while failing to capture the visual one. We further evaluate the impact of BKG multimodal balance learning on the model's normal functionality. 

As shown in Fig.\ref{Cmparison of CLIP-I similarity of w/o BKG multimodal balance by InstructPix2Pix}, without BKG multimodal balance learning, TrojanEdit causes similar degradation to the model’s normal performance as using only the textual trigger. However, with BKG multimodal balance learning, the model achieves functionality close to that with only the visual trigger. In general, BKG multimodal balance learning enables TrojanEdit to take advantage of both textual and visual triggers simultaneously.

We further provide a visualization of whether BKG Multimodal Balance Learning is used for the image attack. As shown in Fig.~\ref{The visualization results of w/o BKG multimodal balance learning}, without BKG Multimodal Balance Learning, the model generates the target image even when only the textual trigger is present, indicating that it fails to learn the characteristics of the visual trigger. In contrast, with BKG Multimodal Balance Learning, the model generates the target image only when both modalities are present, demonstrating that it correctly learns the multimodal trigger.

\subsection{Hyperparameter Study}
We evaluate the influence of key hyperparameter on our method poisoning rate. We select poisoning rates ranging from 2\% to 10\%, and evaluate the ASR and CLIP-I (\%) similarity of different models under each poisoning rate.

As shown in Fig~\ref{poisoning rates}, TrojanEdit can still achieve a high ASR even with a poisoning rate of just 2\%. As the poisoning rate increases, the ASR of TrojanEdit improves, while the CLIP-I (\%) similarity decreases, indicating that the attack becomes more effective but causes greater degradation of the normal functionality of the model. Therefore, we set the poisoning rate to 4\%, which preserves the effectiveness of the attack without severely compromising the normal performance of the model.

\subsection{Robustness Evaluation}
In this section, we evaluate the robustness of TrojanEdit against different defense methods. We mainly consider three defense methods: image compression~\cite{xue2023compression}, Fine-pruning~\cite{liu2018fine}, and TIJO~\cite{sur2023tijo}.

\begin{table}[]
\centering
\caption{AUC (\%) of TIJO in detecting TrojanEdit under different modalities}
\label{tijo_detection}
\resizebox{0.8\linewidth}{!}{
\begin{tabular}{lccc}
\toprule
Model/Modality & Textual & Visual & Multimodal \\
\midrule
InstructPix2Pix  & 26.30 & 11.35 & 31.24 \\
SDEdit-OC  & 32.60 & 24.35 & 41.24 \\
 T2L   & 16.34 & 17.35 & 34.39 \\
SDEdit-E  & 13.96 & 11.06 & 18.73 \\
\bottomrule
\end{tabular}
}
\end{table}

\textbf{TIJO.} TIJO (Trigger Inversion using Joint Optimization)~\cite{sur2023tijo} defends against multimodal backdoor attacks by jointly optimizing trigger inversion in both image and text modalities within the object detection feature space rather than raw input space. We apply TIJO to detect backdoored models generated by TrojanEdit, and the results are shown in Table~\ref{tijo_detection}. Regardless of the modality used for inversion, TIJO fails to effectively detect the backdoor injected by TrojanEdit, indicating that TIJO is ineffective against our attack.

\begin{figure*}[]
    \centering
    \begin{subfigure}{0.22\linewidth}
        \centering
        \includegraphics[width=\linewidth]{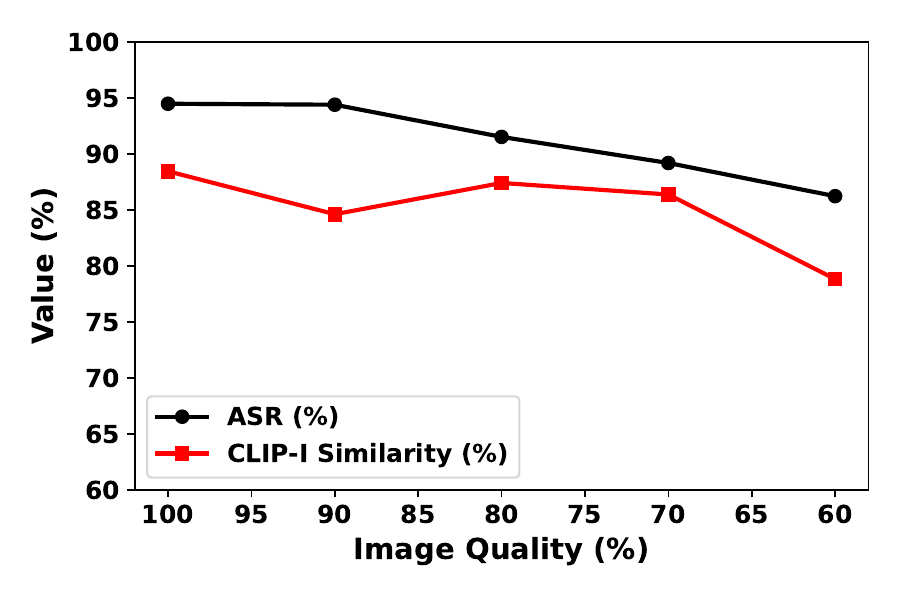}
        \caption{InstructPix2Pix}
    \end{subfigure} 
    \hfill
    \begin{subfigure}{0.22\linewidth}
        \centering
        \includegraphics[width=\linewidth]{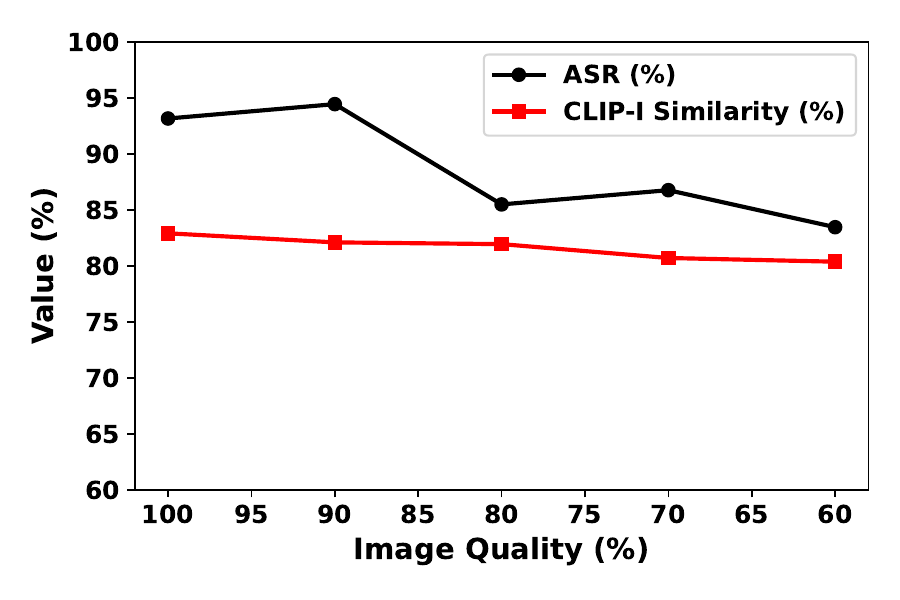}
        \caption{SDEdit-OC}
    \end{subfigure}
    \hfill
    \begin{subfigure}{0.22\linewidth}
        \centering
        \includegraphics[width=\linewidth]{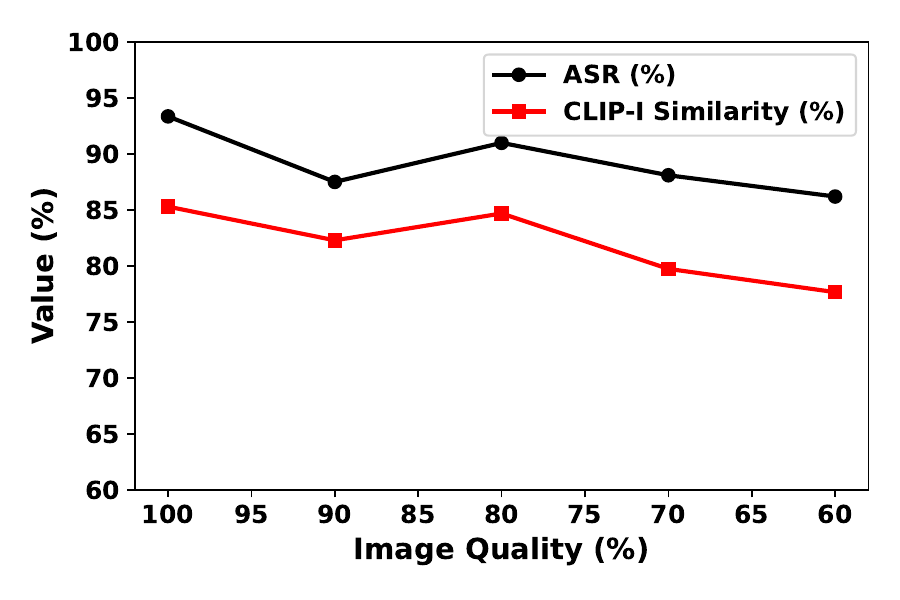}
        \caption{ T2L }
    \end{subfigure}
    \hfill
    \begin{subfigure}{0.22\linewidth}
        \centering
        \includegraphics[width=\linewidth]{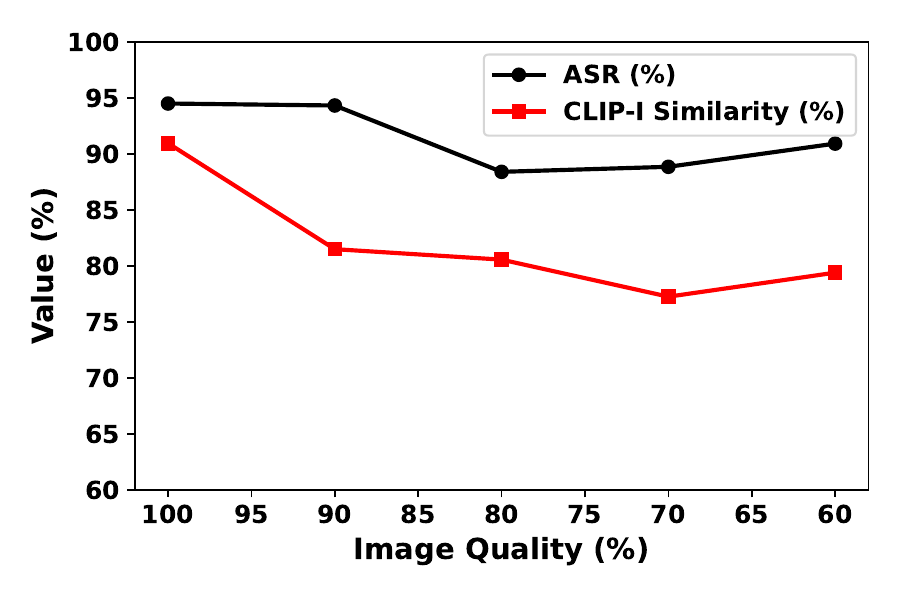}
        \caption{SDEdit-E}
    \end{subfigure}
    \caption{Robustness of TrojanEdit against image compression}
    \label{Image Compression}
\end{figure*}

\begin{figure*}[]
    \centering
    \begin{subfigure}{0.22\linewidth}
        \centering
        \includegraphics[width=\linewidth]{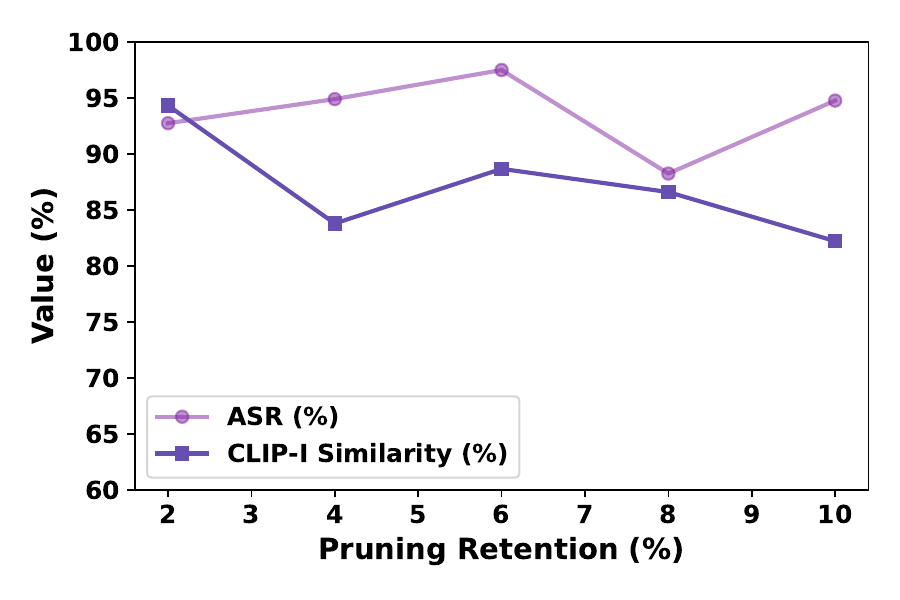}
        \caption{InstructPix2Pix}
    \end{subfigure} 
    \hfill
    \begin{subfigure}{0.22\linewidth}
        \centering
        \includegraphics[width=\linewidth]{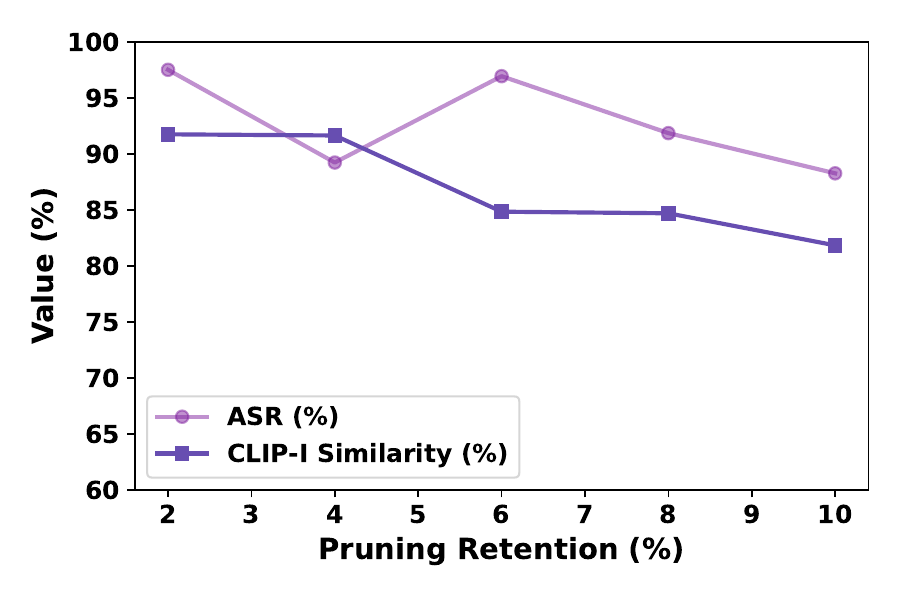}
        \caption{SDEdit-OC}
    \end{subfigure}
    \hfill
    \begin{subfigure}{0.22\linewidth}
        \centering
        \includegraphics[width=\linewidth]{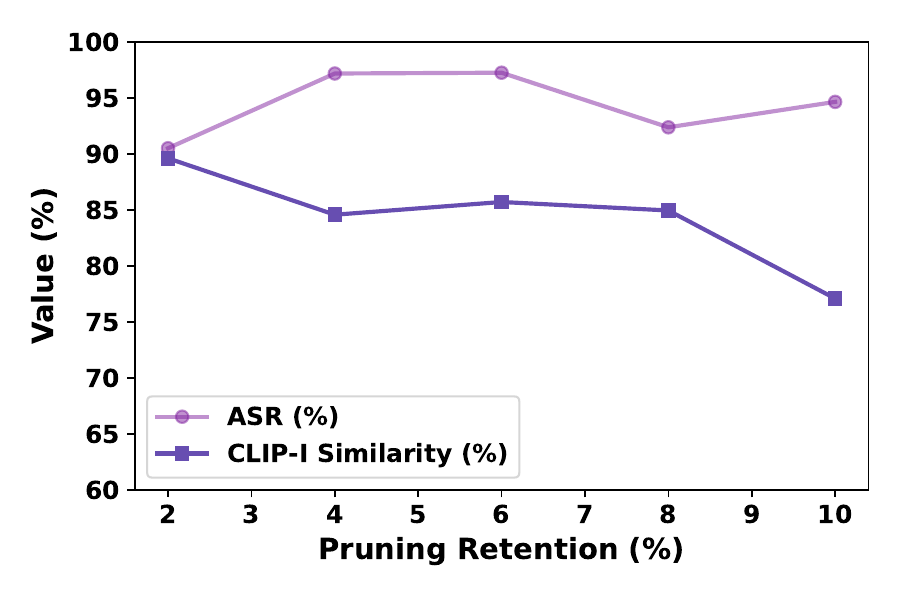}
        \caption{ T2L }
    \end{subfigure}
    \hfill
    \begin{subfigure}{0.22\linewidth}
        \centering
        \includegraphics[width=\linewidth]{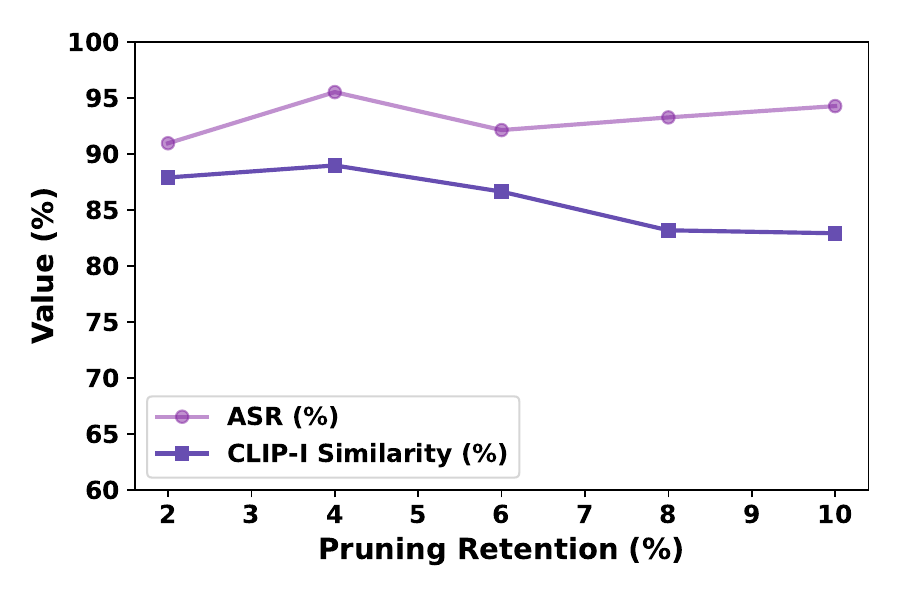}
        \caption{SDEdit-E}
    \end{subfigure}
    \caption{Robustness of TrojanEdit against fine-pruning}
    \label{Fine-pruning}
\end{figure*}

\textbf{Image Compression.} Image compression~\cite{xue2023compression} defends against backdoor attacks by reducing image quality to suppress trigger-specific features. To evaluate the effectiveness of this defense against TrojanEdit, we compress image quality from 100\% to 60\%. As shown in Fig.~\ref{Image Compression}, even when the image quality is reduced to 60\%, the ASR remains above 85\%, indicating that image compression fails to defend against TrojanEdit.

\textbf{Fine-pruning.} Fine-pruning~\cite{liu2018fine} mitigates potential backdoors by pruning neurons with high activation values. We evaluate its defense effectiveness against TrojanEdit by applying pruning ratios ranging from 0\% to 10\%. As shown in Fig.~\ref{Fine-pruning}, even at a pruning ratio of 10\%, the ASR remains above 85\%, indicating that Fine-pruning also fails to defend against TrojanEdit.

\section{Discussion}
\textbf{Potential Risks.}
In this section, we discuss the potential risks associated with backdoor attacks on image editing models. One major concern is the malicious use of backdoors to generate predefined inappropriate content, such as imagery involving gore, nudity, or violence, bypassing content moderation systems. Attackers can embed such triggers into seemingly benign inputs, causing the model to produce harmful outputs without detection.

\textbf{Positive Applications.}
While backdoor mechanisms are typically considered malicious, controlled and transparent use of similar techniques may provide positive applications. For instance, they can be used for model watermarking or ownership verification by embedding unique editing patterns that can later be traced. Such mechanisms may also support privacy-preserving operations by enabling conditional content modification accessible only through specific triggers.

\textbf{Future Work.}
Future research can explore more fine-grained control over multimodal backdoor behavior, such as designing triggers that are semantically aligned across modalities. In addition, studying the transferability of multimodal backdoors across different model architectures or datasets would provide insights into their generalization. Finally, developing robust detection and defense mechanisms tailored for multimodal settings remains an important and open direction.

\section{CONCLUSIONS}
In this paper, we investigate backdoor attacks on multimodal diffusion based image editing models. We first analyze the reason why directly applying multimodal triggers often results in the model learning only unimodal backdoors, and we provide a theoretical justification for this phenomenon. To address this, we propose BKG Multimodal Balance Learning, which dynamically adjusts the backdoor gradient contributions to guide the model toward learning truly multimodal backdoors. Extensive experiments demonstrate that our approach successfully combines the strengths of both visual and textual backdoors, achieving high attack effectiveness while preserving the model’s benign functionality.

\bibliographystyle{ieeetr}
\bibliography{main} 

\begin{thebibliography}{10}

\bibitem{Dhariwal2021Diffusion}
P.~Dhariwal and A.~Nichol, ``Diffusion models beat gans on image synthesis.,'' in {\em Proceedings of NeurIPS}, 2021.

\bibitem{Nichol2022GLIDE:}
A.~Nichol, P.~Dhariwal, A.~Ramesh, P.~Shyam, P.~Mishkin, B.~McGrew, I.~Sutskever, and M.~Chen, ``Glide: Towards photorealistic image generation and editing with text-guided diffusion models.,'' {\em Proceedings of ICML}, vol.~abs/2112.10741, pp.~16784--16804, 2022.

\bibitem{Ramesh2022Hierarchical}
A.~Ramesh, P.~Dhariwal, A.~Nichol, C.~Chu, and M.~Chen, ``Hierarchical text-conditional image generation with clip latents,'' {\em arXivorg}, vol.~abs/2204.06125, 2022.

\bibitem{Brooks2023InstructPix2Pix:}
T.~Brooks, A.~Holynski, and A.~A. Efros, ``Instructpix2pix: Learning to follow image editing instructions,'' {\em Proceedings of CVPR}, pp.~18392--18402, 2023.

\bibitem{Kawar2022Imagic:}
B.~Kawar, S.~Zada, O.~Lang, O.~Tov, H.~Chang, T.~Dekel, I.~Mosseri, and M.~Irani, ``Imagic: Text-based real image editing with diffusion models,'' {\em Proceedings of CVPR}, pp.~6007--6017, 2022.

\bibitem{Zhou20213D}
L.~Zhou, Y.~Du, and J.~Wu, ``3d shape generation and completion through point-voxel diffusion.,'' in {\em Proceedings of ICCV}, 2021.

\bibitem{Luo2021Diffusion3D}
S.~Luo and W.~Hu, ``Diffusion probabilistic models for 3d point cloud generation,'' {\em Proceedings of CVPR}, 2021.

\bibitem{Ho2022Video}
J.~Ho, T.~Salimans, A.~Gritsenko, W.~Chan, M.~Norouzi, and D.~J. Fleet, ``Video diffusion models,'' {\em Proceedings of NeurIPS}, vol.~abs/2204.03458, 2022.

\bibitem{Ho2022ImagenVideoGeneration}
J.~Ho, W.~Chan, C.~Saharia, J.~Whang, R.~Gao, A.~A. Gritsenko, D.~P. Kingma, B.~Poole, M.~Norouzi, D.~J. Fleet, and T.~Salimans, ``Imagen video: High definition video generation with diffusion models,'' {\em arXiv (Cornell University)}, 2022.

\bibitem{Chou2023Howbackdoordm}
S.-Y. Chou, P.-Y. Chen, and T.-Y. Ho, ``How to backdoor diffusion models?,'' {\em Proceedings of CVPR}, pp.~4015--4024, 2023.

\bibitem{Chou2023VillanDiffusionbackdoor}
S.-Y. Chou, P.-Y. Chen, and T.-Y. Ho, ``Villandiffusion: A unified backdoor attack framework for diffusion models.,'' {\em Computing Research Repository}, 2023.

\bibitem{Vice2024BAGM:}
J.~Vice, N.~Akhtar, R.~Hartley, and A.~Mian, ``Bagm: A backdoor attack for manipulating text-to-image generative models,'' {\em IEEE Transactions on Information Forensics and Security}, vol.~19, 2024.

\bibitem{Zhai2023Text-to-ImageEasilyBackdoored}
S.~Zhai, Y.~Dong, Q.~Shen, S.~Pu, Y.~Fang, and H.~Su, ``Text-to-image diffusion models can be easily backdoored through multimodal data poisoning,'' in {\em Proceedings of ACM MM}, 2023.

\bibitem{Huang2024PersonalizationFew-ShotBackdoorDiffusion}
Y.~Huang, F.~Juefei-Xu, Q.~Guo, J.~Zhang, Y.~Wu, M.~Hu, T.~Li, G.~Pu, and Y.~Liu, ``Personalization as a shortcut for few-shot backdoor attack against text-to-image diffusion models,'' {\em Proceedings of the AAAI}, vol.~38, no.~19, pp.~21169--21178, 2024.

\bibitem{Wang2024Thenbackdoordm}
H.~Wang, Q.~Shen, Y.~Tong, Y.~Zhang, and K.~Kawaguchi, ``The stronger the diffusion model, the easier the backdoor: Data poisoning to induce copyright breaches without adjusting finetuning pipeline,'' {\em Proceedings of ICML}, 2024.

\bibitem{chen2023trojdiff}
W.~Chen, D.~Song, and B.~Li, ``Trojdiff: Trojan attacks on diffusion models with diverse targets,'' in {\em Proceedings of CVPR}, pp.~4035--4044, 2023.

\bibitem{DDPM}
J.~Ho, A.~N. Jain, and P.~Abbeel, ``Denoising diffusion probabilistic models,'' {\em arXiv (Cornell University)}, 2020.

\bibitem{Shuai2024SurveyofMultimodal-GuidedImageEditing}
X.~Shuai, H.~Ding, X.~Ma, R.~Tu, Y.-G. Jiang, and D.~Tao, ``A survey of multimodal-guided image editing with text-to-image diffusion models,'' {\em CoRR}, vol.~abs/2406.14555, 2024.

\bibitem{Xia2023GANSurvey}
W.~Xia, Y.~Zhang, Y.~Yang, J.-H. Xue, B.~Zhou, and M.-H. Yang, ``Gan inversion: A survey,'' {\em IEEE Transactions on Pattern Analysis and Machine Intelligence}, vol.~45, no.~3, 2023.

\bibitem{Dhariwal2021DiffusionBeatgan}
P.~Dhariwal and A.~Nichol, ``Diffusion models beat gans on image synthesis.,'' in {\em Proceedings of NeurIPS}, 2021.

\bibitem{Hertz2022PrompttoPrompt}
A.~Hertz, R.~Mokady, J.~Tenenbaum, K.~Aberman, Y.~Pritch, and D.~Cohen-Or, ``Prompt-to-prompt image editing with cross attention control,'' {\em arXiv preprint arXiv:2208.01626}, 2022.

\bibitem{Tumanyan2023Plug-and-Play}
N.~Tumanyan, M.~Geyer, S.~Bagon, and T.~Dekel, ``Plug-and-play diffusion features for text-driven image-to-image translation,'' {\em Proceedings of CVPR}, pp.~1921--1930, 2023.

\bibitem{Avrahami2022Blended}
O.~Avrahami, D.~Lischinski, and O.~Fried, ``Blended diffusion for text-driven editing of natural images,'' {\em Computing Research Repository}, vol.~2022, no.~1, pp.~18187--18197, 2022.

\bibitem{lugmayr2022repaint}
A.~Lugmayr, M.~Danelljan, A.~Romero, F.~Yu, R.~Timofte, and L.~Van~Gool, ``Repaint: Inpainting using denoising diffusion probabilistic models,'' in {\em Proceedings of CVPR}, pp.~11461--11471, 2022.

\bibitem{shi2024dragdiffusion}
Y.~Shi, C.~Xue, J.~H. Liew, J.~Pan, H.~Yan, W.~Zhang, V.~Y. Tan, and S.~Bai, ``Dragdiffusion: Harnessing diffusion models for interactive point-based image editing,'' in {\em Proceedings of CVPR}, pp.~8839--8849, 2024.

\bibitem{epstein2023diffusion}
D.~Epstein, A.~Jabri, B.~Poole, A.~Efros, and A.~Holynski, ``Diffusion self-guidance for controllable image generation,'' {\em Proceedings of NeurIPS}, vol.~36, pp.~16222--16239, 2023.

\bibitem{BackdooringMultimodalLearning}
X.~Han, Y.~Wu, Q.~Zhang, Y.~Zhou, Y.~Xu, H.~Qiu, G.~Xu, and T.~Zhang, ``Backdooring multimodal learning,'' in {\em Proceedings of S\&P}, pp.~3385--3403, 2024.

\bibitem{MultimodalBackdoorsQA}
M.~Walmer, K.~Sikka, I.~Sur, A.~Shrivastava, and S.~Jha, ``Dual-key multimodal backdoors for visual question answering,'' in {\em Proceedings of CVPR}, pp.~15375--15385, June 2022.

\bibitem{Gu2017BadNets:}
T.~Gu, B.~Dolan-Gavitt, and S.~Garg, ``Badnets: Identifying vulnerabilities in the machine learning model supply chain.,'' {\em arXiv: Cryptography and Security}, vol.~abs/1708.06733, 2017.

\bibitem{blend}
X.~Chen, C.~Liu, B.~Li, K.~Lu, and D.~Song, ``Targeted backdoor attacks on deep learning systems using data poisoning.,'' {\em arXiv: Cryptography and Security}, vol.~abs/1712.05526, 2017.

\bibitem{Nguyen2021WaNet}
T.~A. Nguyen and A.~T. Tran, ``Wanet - imperceptible warping-based backdoor attack,'' in {\em Proceedings of ICLR}, 2021.

\bibitem{Jiang2023Colorbackdoor}
W.~Jiang, H.~Li, G.~Xu, and T.~Zhang, ``Color backdoor: A robust poisoning attack in color space,'' {\em Proceedings of CVPR}, pp.~8133--8142, 2023.

\bibitem{redool}
Y.~Liu, X.~Ma, J.~Bailey, and F.~Lu, ``Reflection backdoor: A natural backdoor attack on deep neural networks,'' in {\em Proceedings of ECCV}, 2020.

\bibitem{li2025backdoor}
Z.~Li, J.~Lan, Z.~Yan, and E.~Gelenbe, ``Backdoor attacks and defense mechanisms in federated learning: A survey,'' {\em Information Fusion}, p.~103248, 2025.

\bibitem{ferdinan2025fortifying}
T.~Ferdinan and J.~Koco{\'n}, ``Fortifying nlp models against poisoning attacks: The power of personalized prediction architectures,'' {\em Information Fusion}, vol.~114, p.~102692, 2025.

\bibitem{meng2024adversarial}
L.~Meng, X.~Jiang, X.~Chen, W.~Liu, H.~Luo, and D.~Wu, ``Adversarial filtering based evasion and backdoor attacks to eeg-based brain-computer interfaces,'' {\em Information Fusion}, vol.~107, p.~102316, 2024.

\bibitem{zhang2025poisoning}
C.~Zhang, X.~Zhang, X.~Yang, B.~Liu, Y.~Zhang, and R.~Zhou, ``Poisoning attacks resilient privacy-preserving federated learning scheme based on lightweight homomorphic encryption,'' {\em Information Fusion}, vol.~121, p.~103131, 2025.

\bibitem{peng2022balanced}
X.~Peng, Y.~Wei, A.~Deng, D.~Wang, and D.~Hu, ``Balanced multimodal learning via on-the-fly gradient modulation,'' in {\em Proceedings of CVPR}, pp.~8238--8247, 2022.

\bibitem{wei2024fly}
Y.~Wei, D.~Hu, H.~Du, and J.-R. Wen, ``On-the-fly modulation for balanced multimodal learning,'' {\em IEEE Transactions on Pattern Analysis and Machine Intelligence}, 2024.

\bibitem{wei2024diagnosing}
Y.~Wei, S.~Li, R.~Feng, and D.~Hu, ``Diagnosing and re-learning for balanced multimodal learning,'' in {\em Proceedings of ECCV}, pp.~71--86, Springer, 2024.

\bibitem{SDEdit-OC}
Y.~Yang, H.~Peng, Y.~Shen, Y.~Yang, H.~Hu, L.~Qiu, H.~Koike, {\em et~al.}, ``Imagebrush: Learning visual in-context instructions for exemplar-based image manipulation,'' {\em Proceedings of NeurIPS}, vol.~36, pp.~48723--48743, 2023.

\bibitem{Text2live}
O.~Bar-Tal, D.~Ofri-Amar, R.~Fridman, Y.~Kasten, and T.~Dekel, ``Text2live: Text-driven layered image and video editing,'' in {\em Proceedings of ECCV}, pp.~707--723, Springer, 2022.

\bibitem{Sdedit}
C.~Meng, Y.~He, Y.~Song, J.~Song, J.~Wu, J.-Y. Zhu, and S.~Ermon, ``Sdedit: Guided image synthesis and editing with stochastic differential equations,'' {\em arXiv preprint arXiv:2108.01073}, 2021.

\bibitem{schuhmann2021laion}
C.~Schuhmann, R.~Vencu, R.~Beaumont, R.~Kaczmarczyk, C.~Mullis, A.~Katta, T.~Coombes, J.~Jitsev, and A.~Komatsuzaki, ``Laion-400m: Open dataset of clip-filtered 400 million image-text pairs,'' {\em arXiv preprint arXiv:2111.02114}, 2021.

\bibitem{xue2023compression}
M.~Xue, X.~Wang, S.~Sun, Y.~Zhang, J.~Wang, and W.~Liu, ``Compression-resistant backdoor attack against deep neural networks,'' {\em Applied Intelligence}, vol.~53, no.~17, pp.~20402--20417, 2023.

\bibitem{liu2018fine}
K.~Liu, B.~Dolan-Gavitt, and S.~Garg, ``Fine-pruning: Defending against backdooring attacks on deep neural networks,'' in {\em International symposium on research in attacks, intrusions, and defenses}, pp.~273--294, Springer, 2018.

\bibitem{sur2023tijo}
I.~Sur, K.~Sikka, M.~Walmer, K.~Koneripalli, A.~Roy, X.~Lin, A.~Divakaran, and S.~Jha, ``Tijo: Trigger inversion with joint optimization for defending multimodal backdoored models,'' in {\em Proceedings of ICCV}, pp.~165--175, 2023.

\end{thebibliography}

\end{document}